\documentclass[a4paper,11pt]{article}
\usepackage{jcappub} 
\usepackage{lineno}
\usepackage{xcolor}
\usepackage{float}
\usepackage{tikz}
\usepackage{amsmath}
\usetikzlibrary{positioning, calc, fit}
\usepackage{mathdots}
\usepackage{cleveref}
\usepackage{makecell}
\usepackage{multirow}
\usepackage{tablefootnote}
\usepackage{array}
\usepackage{textcomp}
\usepackage{multirow}

\usepackage{booktabs}
\usepackage{silence}
\definecolor{myblue}{HTML}{1b5f6f}
\definecolor{orange}{HTML}{E69F00}
\definecolor{lightblue}{HTML}{56B4E9}
\definecolor{darkorange}{HTML}{D55E00}
\usepackage{booktabs}
\usepackage{pdflscape}  
\usetikzlibrary{arrows.meta, positioning, calc}

\usetikzlibrary{positioning,calc,shadows,shapes.geometric,arrows.meta,decorations.pathreplacing,fit}
\usepackage{amsmath}
\usepackage{siunitx}
\usepackage{xcolor}

\definecolor{romancolor}{HTML}{1F78B4}
\definecolor{accent}{HTML}{33A02C}
\definecolor{muted}{HTML}{6A6A6A}
\definecolor{panel}{HTML}{F7FBFF}
\definecolor{boxbg}{HTML}{FFFFFF}
\tikzset{
box/.style = {rectangle, rounded corners=6pt, draw=romancolor!85, very thick, fill=boxbg, drop shadow, inner sep=8pt, text width=5.2cm, align=center, font=\sffamily\small},
smallbox/.style = {rectangle, rounded corners=4pt, draw=muted!70, fill=panel, inner sep=6pt, text width=3.6cm, align=center, font=\sffamily\footnotesize},
process/.style = {ellipse, draw=accent!85, very thick, fill=boxbg, inner sep=6pt, font=\sffamily\small},
data/.style = {cylinder, shape border rotate=90, draw=romancolor!85, very thick, fill=panel, inner ysep=6pt, minimum height=1.2cm, font=\sffamily\small},
arrow/.style = {-{Stealth[length=7pt,width=6pt]}, line width=1pt, draw=black!70},
eq/.style = {font=\small\ttfamily, text=muted}
}
\usepackage{diagbox} 

\usepackage{enumerate}

\usepackage[title]{appendix}

\title{Modeling Redshift Uncertainties in Roman Weak Lensing Cosmology}

\author[a]{Diogo H. F. de Souza,}
\author[b]{Boyan Yin,}
\author[c]{Tim Eifler,}
\author[d]{Vivian Miranda,}
\author[e]{Chun-Hao To,}
\author[f]{Brett H. Andrews,}
\author[a]{Katarina Markovič,}
\author[a]{Eric Huff,}
\author[b]{Michael A. Troxel,}
\author[a]{Olivier Doré}

\affiliation[a]{Jet Propulsion Laboratory, California Institute of Technology, 4800 Oak Grove Drive, Pasadena, CA 91109, USA}
\affiliation[b]{Department of Physics, Duke University Durham, NC 27708, USA}
\affiliation[c]{Department of Astronomy/Steward Observatory, University of Arizona, 933 North Cherry Avenue, Tucson, AZ 85721-0065, USA}
\affiliation[d]{C.N. Yang Institute for Theoretical Physics, Stony Brook University, NY 11794, USA}
\affiliation[e]{Department of Astronomy and Astrophysics, University of Chicago, Chicago, IL 60637, USA}
\affiliation[f]{Department of Physics and Astronomy, University of Pittsburgh, Pittsburgh, PA 15260, USA}
\emailAdd{souzadio@jpl.nasa.gov}

\abstract{
Cosmological constraints using weak gravitational lensing measurements from the Roman Space Telescope will require a powerful method for modelling uncertainties in the galaxy redshift distribution. In this work, we use an optimized version of the principal component analysis (PCA) to model uncertainties in the full shape of the redshift distributions, a method proposed by \cite{pca_method} and recently used in the Dark Energy Survey Y6 analysis. Here, we implement this new approach within the Roman High Latitude Imaging Survey (HLIS) Cosmology Project Infrastructure Team (PIT) pipeline, namely Cobaya-Cosmolike Joint Architecture (\texttt{CoCoA}). To validate the PCA in mitigating biases on cosmological parameters, $S_8$ and $\Omega_m$, we use a set of redshift distributions from \texttt{Cardinal} generated for a variety of Roman configurations. Overall, when the simulated cosmic shear data vector is not strongly miscalibrated relative to the fiducial one, both the mean-shift and the PCA-based approaches produce consistent cosmological constraints when marginalizing over nuisance parameters. For mild to strong miscalibration, including additional PCs progressively mitigates biases in $S_8$ and $\Omega_m$, and can achieve comparable performance with fewer parameters than the nine tomographic-bin mean-shift model.

}

\begin{document}
\maketitle

\section{Introduction}\label{sec:introduction}

Cosmic shear statistics describe the tiny distortions in the shapes of distant background galaxies. These distortions are caused by the weak gravitational lensing of light as it passes through the intervening matter distribution between us and the ensemble of these source galaxies. Cosmic shear is a powerful and complementary probe of the late universe in the era of precision cosmology \cite{wl_precision_cosmology} when combined with galaxy clustering and galaxy-galaxy lensing, forming the so-called $3\times2$pt analysis \cite{obs_probes_cosmic_accel,comprehensive_twopoint_wl_survey,weak_gravitational_lensing}. However constraints on the matter density $\Omega_m$ and the amplitude of mass fluctuations $\sigma_8$ from the cosmic shear signal are affected by several systematic errors such as intrinsic alignment \cite{intrinsic_extrinsic_alignment,intrinsic_alignment_lrg,intrinsic_alignments_wigglez}, anisotropic point spread function \cite{optimal_measurements_for_wl}, baryonic feedback \cite{baryon_pca_wl_1,baryon_pca_wl_2}, and uncertainties in the redshift distribution \cite{catastrophic_photoz}, which is the subject of this work. Partitioning the galaxy sample into redshift bins improves the measurement of cosmological parameters \cite{tomography_wl}. Different redshifts contribute differently to the lensing signal - light from high-z galaxies encounters more structure along the line of sight and thus has a stronger lensing signal. However, \cite{tomography_wl} points out important caveats that one should consider for optimal extraction of cosmological information from binning the redshift distribution of the galaxies samples, such as the survey depth and the accuracy of the photometric redshift technique. A forecast analysis for next-generation surveys of the impact of imperfect tomographic photometric redshift determination on dynamical dark energy was carried out by \cite{effects_of_photoz_u_wl}. They demonstrated the degeneracies between photometric redshift and dark energy parameters, as well as the resulting degradation in the constraining power on dark energy.

The effects of photo-z uncertainties on tomographic weak lensing cosmology were extensively explored in the Dark Energy Survey (DES) Y1 and Y3 analyses by performing a number of robustness to modeling tests. The redshift distribution for sources galaxies in the weak lensing analysis of the DES Y1 is fully discussed in \cite{des_y1_zdistri_of_wl} and \cite{des_y1_methods_systematics_characterizations,des_y1_calibration_wl} for the clustering-based method. Broadly speaking, this procedure consists of obtaining initial estimates of the lensing-weighted redshift distribution by assigning galaxies to four redshift bins using the Bayesian photometric redshift (BPZ) approach of \cite{bpz}, constituting the DES Y1 fiducial analysis. However, several effects contribute to the total uncertainty budget when estimating the redshift distributions, and their impact is commonly compressed into shifts of the mean redshift distribution, $n_i(z) \rightarrow n_i(z - \Delta z_i)$, parameterized by $\Delta z_i$, where $i$ denotes the tomographic bin. The constraints of cosmological parameters is performed by marginalizing over the nuisance photo-z parameters, i.e., the $\Delta z_i$'s. Beyond the mean-shift parameters, the forecast analysis of \cite{effects_of_photoz_u_wl} also considered a Gaussian scatter to describe the photometric redshift distributions. 

Although the inclusion of two sets of parameters-the mean shift and the Gaussian scatter-improves the realism of modeling imperfect photometric redshift distributions, in real analyses one must also consider the associated computational cost of introducing additional nuisance parameters. There are several ways to design tests to assess whether photo-z parameters beyond the mean-shift are required to correct imperfections in $n_i(z)$ from the true redshift distribution. The investigation of this question is closely tied to the current survey precision in measuring photometric redshift distributions. For example, the DES-Y1 analysis of cosmological constraints from cosmic shear presented in \cite{des_y1_constraints_from_cosmic_shear}, considered both the BPZ redshift distribution and the resampled high-precision COSMOS2015 \cite{cosmos2015} 
which exhibit different shapes in the redshift distribution, particularly in the second tomographic bin (see, e.g., Figure~2 of \cite{des_y1_constraints_from_cosmic_shear}). At the DES Y1 statistical precision, the difference in the shape of $n_i(z)$ between BPZ and COSMOS is a subdominant contributor to the uncertainties on cosmic shear and does not significantly impact the inference of the cosmological parameters (see, e.g.,~Figures 9 and 13 of \cite{des_y1_constraints_from_cosmic_shear} for the 1D and 2D constraints for $S_8$ and $S_8-\Omega_m$, respectively). The tomographic weak lensing measurement of \cite{des_y1_constraints_from_cosmic_shear} for the $\Lambda$CDM model reported a $3.5\%$ fractional uncertainty on $\sigma_8(\Omega_m/0.3)^{0.5}=0.782\pm{0.027}$ at $68\%$ CL (for the DES Y1 $3\times2$pt analysis see \cite{des_y1_constraints_from_clustering_and_wl}). 

From the DES Y3 Gold photometric data set \cite{des_y3_gold_photometric}, a weak-lensing source galaxy sample was selected \cite{wl_shape_catalog} and divided into four tomographic bins. The redshift distributions of these bins were calibrated using three independent likelihood-based methods-the Self-Organizing Map $p(z)$ (SOMPZ) \cite{des_y3_som}, clustering redshifts (WZ) \cite{des_y3_wz}, and shear ratios (SR) \cite{des_y3_sr}. Together, these complementary techniques yielded consistent and robust constraints on the mean redshift of each bin, achieving an uncertainty of $\sigma_{\langle z\rangle}\approx0.01$ \cite{des_y3_calibration_wl}. The resulting calibrated redshift distributions played a key role in assessing the robustness of the DES Y3 cosmic shear cosmological constraints on both modeling choices and data calibration \cite{des_y3_cosmic_shear_robustness,des_y3_cosmic_shear_robustness_to_calib} (see also \cite{des_y3_cosmology_from_clustering_and_wl} for the full $3\times2$pt analysis). In these studies, the fiducial photometric redshift model parameterizes statistical and systematic calibration uncertainties through the mean redshift shifts, $\Delta z_i$. The analysis in \cite{des_y3_cosmic_shear_robustness_to_calib}, however, extended this approach by incorporating the full uncertainty in the shape of the redshift distributions using the Hyperrank formalism \cite{des_y3_hyperrank}. The one- and two-dimensional posterior constraints on $S_8$ and $S_8$-$\Omega_m$ shown in Figures 7 and 10 of \cite{des_y3_cosmic_shear_robustness_to_calib} demonstrated that the full-shape Hyperrank model yielded results consistent within $1\sigma$ of the fiducial DES Y3 analysis.

The Hyperrank approach consists of marginalizing over the range of uncertainty spanned by the ensemble of plausible redshift distributions constructed using the SOMPZ method from the \texttt{Buzzard} simulated photometry catalog for DES Y3. These redshift distributions are ranked according to a descriptive statistic (e.g., the mean redshift) that a priori correlates with the cosmological parameters of interest, such as $S_8$. The summary statistic is then used to encode each realization in a optimal relative position of a uniform grid embedded within a unit hypercube. These unit coordinates are sampled in a Markov chain during likelihood inference, and the continuous hyper-parameter value is mapped to a discrete index, which can be decoded to recover the corresponding redshift distribution. However, the performance of Hyperrank depends on balancing the number of redshift realizations, the dimensionality of the hypercube, and the choice of descriptive statistics. Increasing the dimensionality can in principle yield a smoother representation of the posterior, but for a fixed number of realizations this reduces the resolution of the grid and introduces larger spacing between samples. This, in turn, can lead to noisier and less continuous posteriors in the hyper-parameter space, which may hinder sampling efficiency. We refer the reader to \cite{des_y3_hyperrank} for further details on Hyperrank. 

An alternative way to marginalize over the shape uncertainty of the redshift distribution is an optimized version of the principal component approach first developed by \cite{pca_method} and used in this work. In this PCA method, a compression matrix $\mathbf{E}$ is constructed to encode the variations in the redshift distributions into a lower-dimensional set of variables $\mathbf{u}$. This effectively compresses the ensemble into the most informative degrees of freedom, freeing the likelihood of less significant variations in the redshift distribution. A decoder matrix $\mathbf{D}$ is used to build basis functions (we use the terms ``PCs'' and ``modes'' interchangeably) to recover an approximation for the redshift distribution represented by the most informative PCs. This technique was used in the DES Y6 analysis for the redshift calibration of sources galaxies \cite{des_y6_pca_redshift_calib_source} and lens galaxies \cite{des_y6_pca_redshift_calib_lens}. Overall, this mode-projection method has minimal impact on the cosmological parameters given the statistical power of DES Y6 and the robust constraining power on the full shape of the redshift distributions from SOMPZ and WZ. However, continuing investigation of photo-z uncertainties is needed for next-generation Stage IV surveys, such as the Nancy Grace Roman Space Telescope (Roman) \cite{wfirst_essential_cosmology,wfirst_annual_report,wfirst_100_hubbles,roman_multiprobe,wfirst_snia_forecast,roman_synergies_with_cmb}, the NSF-DOE Vera C.~Rubin Observatory Legacy Survey of Space and Time (LSST) \cite{lsst_ref}, and ESA's Euclid mission \cite{euclid_mission_overview}. This paper applies the mode-projection technique devised by \cite{pca_method} for redshift distributions simulated for Roman.   

This paper is organized as follows: in Section \ref{sec:nz_simulations} we present a description of simulated redshift distributions for various proposed observing scenarios for the Roman High-Latitude Wide Area Survey (HLWAS) with \texttt{Cardinal} simulations. In Section \ref{sec:modeling_pca}, we construct the PCA basis and the optimized mode-projection according to the technique of \cite{pca_method} suitable for the Roman needs. In Section \ref{sec:analysis_methodology} we provide a detailed description of the analysis methodology. In Section~\ref{sec:results}, we present our main results: (i) a consistency check of our numerical PCA implementation in CoCoA against the prescription proposed by \cite{pca_method}; (ii) an assessment of cosmological parameter constraints obtained with the PCA and mean-shift approaches; and (iii) a more detailed investigation of the various Roman observing scenarios under the PCA method. Finally, in Section \ref{sec:conclusion}, we present our summary and conclusion with future perspectives.

\section{Simulations of Roman redshift distributions}\label{sec:nz_simulations}
\subsection{Synthetic sky mock catalogs}

Investigating biases in different cosmological probes arising from unknown or mismodeled systematic effects is crucial for obtaining unbiased cosmological constraints. To quantify these systematics, it is necessary to generate synthetic sky mock catalogs with a high level of realism. To be useful for systematic control, such catalogs must satisfy several requirements. First, the mock catalogs should span a volume significantly larger than that of the target survey so that the uncertainty in the estimated systematic effects remains well below the statistical uncertainty of the data. Second, they must include realistic galaxy populations tailored to the specifications of the survey. Third, efficient mock generation is required to enable the incorporation of new techniques and systematic studies \cite{cardinal_sims}. To address these challenges, a wide range of methods for synthetic catalog generation have been developed. In particular, the galaxy-dark matter halo connection framework \cite{galaxy_halo_connection} provides a physically motivated approach; however, mock generation within this framework remains computationally expensive. Phenomenological models offer a practical alternative to mitigate this bottleneck, especially subhalo abundance matching (SHAM) model, a method for mapping galaxy stellar masses or luminosities onto resolved dark matter (sub)halos \cite{sham_orphan_galaxies,sham_1,sham_2,subhalo_abundance_matching}, on which the \texttt{Buzzard} simulations are based. 

The \texttt{Buzzard} simulations comprise a suite of synthetic sky catalogs that typically include galaxy positions, magnitudes, shapes, photometric errors, and photometric redshift estimates, and was used to support a wide range of DES science analysis \cite{buzzard_flock,buzzard_desy3}. These catalogs are widely used for systematic studies and for testing and validating multiple cosmological probes, such as galaxy clustering and weak lensing analyses. Ultimately, they enable end-to-end validation of the cosmological analysis pipeline, including target selection algorithms such as \texttt{redMaGiC} \cite{redmagic} for luminous red galaxies and \texttt{redMaPPer} \cite{redmapper} for galaxy clusters. Despite their broad utility, the \texttt{Buzzard} simulations exhibit a deficit of \texttt{redMaPPer} clusters, with abundances lower by a factor of three to four compared to DES Y1. This mismatch in the simulated galaxy population within cluster environments constitutes a key shortcoming, as realistic cluster galaxy populations are critical for simulation-based evaluations of optically selected cluster detection and characterization \cite{cardinal_sims}. Specifically, one could see discussions that follows Figures 12 and 13 of \cite{buzzard_flock} that shows the deficit in richness at fixed mass when compared to the DES-Y1. As discussed in \cite{addgals}, this issue in the \texttt{Buzzard} simulations was likely due to the artificial subhalo disruption, which is then inherited by the \texttt{addgals} algorithm via the training process. Attempts to address this issue were made by \cite{sham_orphan_galaxies} by including orphan galaxies in SHAM. However, none of the tested models could fit all three stellar mass bins simultaneously, largely because the parameter controlling the asymptotic maximum circular velocity of the halo, strongly affects SHAM predictions. In addition, the models of color-assignment tested in that work underestimates clustering of low-mass red galaxies in the smallest mass bin, which can reduce the number of galaxies in clusters; see \cite{cardinal_sims}, \cite{sham_orphan_galaxies} and references therein for further details.  

\subsection{Cardinal simulations}

The \texttt{Cardinal}\footnote{\url{https://chunhaoto.com/cardinalsim}}
simulations \cite{cardinal_sims} solves this deficit of cluster galaxies in \texttt{Buzzard} simulations by quantifying and correcting the two main sources of the problem: the artificial subhalo disruption in SHAM and the limitations in the color-assignment scheme. This improved model is then passed to the \texttt{Addgals} algorithm and generate \texttt{Cardinal} catalogs, from LSST bands, constructed from a one-quarter--sky simulation populated with galaxies up to redshift $z = 2.35$ and reaching a limiting magnitude of $m_r = 27$. The galaxy magnitudes, colors and several other galaxy properties are provided through the \texttt{Addgals} technique. In the following, we summarize the approach to assign magnitudes and colors to galaxy; firstly, \texttt{Cardinal} assumes that galaxy SEDs can be decomposed into five KCORRECT spectral templates \cite{kcorrect} with associated coefficients. Therefore, one can map the measured the KCORRECT coefficients from real data and assign then to galaxies in the simulations. From this scheme, one can calculate the observed magnitude by applying band shifts and the observed bandpass without regenerating galaxy SEDs. For the color assignment, the PRIMUS galaxies \cite{primus} are used to infer the probability of being red at $z<0.2$. The \texttt{Cardinal} code classifies a galaxy as red if its rest-frame color K-corrected to $z = 0.1$ is above a brightness-dependent threshold of $0.15-0.03M_r$, where $M_r$ is the absolute magnitude of a galaxy in the r-band.

By updating the subhalo abundance matching framework and introducing a new color assignment scheme (see Appendix K of \cite{cardinal_sims} for a summary of the improvements from \texttt{Buzzard 2.0} to \texttt{Cardinal}), the \texttt{Cardinal} mock catalogs enable a wide range of applications. These include (i) studies of optically selected galaxy clusters to quantify selection biases, (ii) evaluation of small-scale lensing systematics using the latest ray-tracing implementation in \texttt{Calclens} \cite{calclens}, (iii) end-to-end tests of multi-probe cosmological analysis, and (iv) investigations of photometric redshift performance (see \cite{cardinal_sims} for further details on each of these points).
In this work, we focus on the latter application. Specifically, the \texttt{Cardinal} code enables us to study how redshift errors in deep-field surveys affect photometric redshift calibration, and how sample variance due to the limited area of deep fields propagates into redshift uncertainties for wide-field galaxies. By permuting different configurations of the deep and wide tiers, we form the scenarios shown in Fig.~\ref{fig:roman_scenarios}, where we generate ensembles of redshift distributions and then, in the next sections, we show how we evaluate their impact on cosmological parameter constraints, such as $\Omega_m$ and $S_8$. In Fig.~\ref{fig:roman_scenarios} we notice, however, a minor caveat; the \texttt{Cardinal} simulations were not designed to span all wide-deep field permutations, and the D3 configuration was included solely as a targeted stress test of a larger deep field for the baseline wide-field scenarios (W1 and W2), which is why no Scenario W3-D3 combination was generated. A subsample of 100 realizations are shown as fainter lines in Fig.~\ref{fig:roman_scenarios} with the scatter reflecting the uncertainty in their estimation. The mean of the full ensemble is represented by the darker, dashed line, whereas different colors and panels represent different Roman scenarios, as indicated by the inner legend.

\subsection{Photometric redshift inference for Roman using SOM}
We use the Self Organizing Map Photo-z (SOMPZ) framework with LSST bands \cite{sompz_des} to estimate the redshift distribution of the wide tier galaxies for Roman. This method was employed in DES-Y6 for redshift calibration for weak lensing studies, see e.g., Section 4 of \cite{des_y6_pca_redshift_calib_source}. Two SOMs are used inside the framework: The wide tier galaxy are grouped onto the ``wide tier SOM'' based on their photometry. The deep-field galaxies, which span less area but have additional photometry and depth are grouped onto the “deep SOM”. For each deep-field phenotype (i.e., cell in the two dimensional “deep SOM”). The redshift distribution is obtained using an external redshift calibration sample. In our analysis, we assume it is our deep field sample with perfect spectroscopic redshift information. We then connect the deep-field phenotypes to wide-tier detections under the same observing conditions using Cardinal simulation. The wide tier galaxies are further separated into 9 tomographic bins, using an equal-populated approach. In particular, we define our redshift bin edges, and each wide tier SOM cell is assigned to the bin according to the mode of its redshift distribution. These bin edges are iteratively adjusted until each tomographic bin contains an approximately equal number of galaxies.

Each redshift realization represents a possible instance of our redshift distribution under the limitation of our calibration sample. Deep fields, while it greatly improves our redshift calibration by providing additional photometry information to break color-redshift degeneracy, have limited sky area and galaxy therein. Due to the limited sky area, the deep field could not fully represent the fluctuations in the matter density field, with the risk of being in an under- or over-dense region. We measure our sample variance with the same method as \cite{sample_variance_propagation,sompz_des,des_y6_pca_redshift_calib_source}, which sums the large scale modes in the CAMB-generated \cite{camb2} matter power spectrum that are not captured within the deep field area.

On the other hand, the finite number of galaxies in the deep field introduce Poisson fluctuation on the number of observed galaxies. The sample variance and shot noise are propagated to our final redshift realizations using a three-step Dirichlet sampling method \cite{sample_variance_propagation,sompz_des,des_y6_pca_redshift_calib_source}. In this analysis, we do not include the uncertainty due to the photometry zero-point offsets in our deep field and wide tier field, since our goal was to estimate the choice of deep field area and depth on redshift uncertainty. These uncertainty will be included in future analysis.

\subsection{Redshift distribution from wide- and deep-tiers permutations}
The wide- and deep-tier configurations for the Roman HLWAS scenarios are shown in Table \ref{tab:roman_scenarios}. The wide tier is designed to yield about 360 million galaxy shape measurements for weak lensing \cite{rotac_report}, whereas the deep tier is intended to support the calibration of the cosmological analysis, particularly the uncertainties in shape measurements, photometric redshift estimation, and spectroscopic contamination correction. In principle, biases in the inferred cosmological parameters arising from miscalibrated photometric redshifts could be constrained using the deep fields D1234. However, the simulations used in this work predate the ROTAC Report \cite{rotac_report}. For this reason, we do not define a Design Reference Mission (DRM). According to the ROTAC recommendations, the DRM corresponds to a wide-tier configuration closest to W3, with the following updated survey properties: an area of $2415\deg^2$, the same three imaging bands (Y, J, and H), an exposure time of $2 \times 3 \times 107$ s, and a depth of 26.4 mag (AB, $5\sigma$ point source) in the JH bands.

\setlength{\tabcolsep}{3pt}
\begin{table}[h]
\centering
\setlength{\tabcolsep}{2.5pt}
\begin{tabular}{|c|c|c|c|c|c|}
\hline
\multicolumn{6}{|c|}{\textbf{High Latitude Wide Area Survey}} \\
\hline
\textbf{Tier} &\textbf{Magnitude Error}& \textbf{Exposure Times} & \textbf{Filters} & \textbf{Area} & \textbf{Depth (H)} \\
\hline
Wide 1 (W1)        & $0.20$  & $5 \times 140$ s & YJHF & $2000\deg^2$& 24.96 \\
Wide 2 (W2)    & $0.25$  & $5 \times 91$ s  & H    & $2000\deg^2$& 23.93 \\
Wide 3 (W3)  & $0.25$  & $5 \times 91$ s  & YJH  & $2000\deg^2$& 24.31 \\
\hline
Deep 1 (D1) & $0.00$ &  $\infty$  & ZYJHFK  & $20\deg^2$ & $\infty$ \\
Deep 2 (D2) & $0.12$ & $5\times91\times4$ s  & ZYJHFK  & $20\deg^2$ & 24.93 \\
Deep 3 (D3) & $0.12$ & $5\times91\times4$ s  & ZYJHFK  & $40\deg^2$ & 24.93 \\
Deep 4 (D4) & $0.12$ & $5\times91\times4$ s  & ZYJHFK  &  $20\deg^2$ & 25.93 \\
\hline
\end{tabular}
\caption{The Roman HLWAS proposed scenarios for the wide and deep tiers prior to the ROTAC recommendations of April 24, 2025. Depths are given as $5\sigma$ AB for an exponential profile with a half-light radius $r_\mathrm{half}=0.3$ arcsec.}
\label{tab:roman_scenarios}
\end{table}

In Table \ref{tab:roman_scenarios}, we use the shortened versions of the filter names for compactness, but their full Roman filter names and central wavelengths are as follows:  Z (F087), Y (F106), J (F129), H (F158), F (F184), and K (F213). Exposure times (ET) refer to individual exposures of 140 s (or 91 s) using a scan pattern of 5 exposures. The idealized deep field D1 has an infinite exposure time, infinite depth, and zero magnitude error. The deep fields D234 have exposure times $4\times$ longer than M23. The deep tiers D23 and D4 are 1 and 2 magnitudes deeper than W2, respectively.

\begin{itemize}
    \item \textbf{Wide Tier 1}: This wide tier corresponds to the design reference mission initially proposed in SRD RST-SYS-REQ-0020, Revision D. However, for weak lensing measurements, the ROTAC recommended dropping the F184 photometric band due to its minor impact on photo-z estimation at $z<3$. Therefore, this makes W3, instead of W1, the closest wide tier scenario to the nominal DRM in the ROTAC report, at least in terms of filters choice. Every Roman HLWAS tier in Table \ref{tab:roman_scenarios} assumes a signal-to-noise ratio (S/N) of 5, then we set a reference for the magnitude error $\sigma_\mathrm{mag}\propto1/\mathrm{S/N}=0.2$ and scale $\sigma_\mathrm{mag}$ for the other tiers according to its exposure time given that $\mathrm{S/N}\propto\sqrt{t_\mathrm{exp}}$ where $t_\mathrm{exp}$ is the exposure time. Finally, W1 has a limiting magnitude in the H band of 24.96 mag (5$\sigma$ AB, exponential profile with half-light radius $r_\mathrm{half}=0.3$ arcsec) over $2000\deg^2$.

    \item \textbf{Wide Tier 2}: This wide tier is similar to the in-guide wide tier in terms of filter choice - both use the same single band (H, F158). However, the ROTAC report expands the wide tier area to $2700 \deg^2$, while here we use $2000 \deg^2$. The magnitude error increases relative to W1 due to shorter exposure times.
    
    \item \textbf{Wide Tier 3}: This is the wide tier most similar to the ROTAC recommendation in terms of the three bands (YJH) of the near infrared imaging. W3 has the same $\sigma_\mathrm{mag}$, $t_\mathrm{exp}$, and area as W2 but is 0.38 mag deeper.
\end{itemize}

The deep tiers D1234 in Table \ref{tab:roman_scenarios} incorporate a deep survey component aimed at improving the calibration of cosmological measurements \cite{rotac_report} with locations that overlap with the LSST Deep Drilling Fields (DDFs). Every deep field has the same imaging filters ZYJHFK, which adds the Z, F, and K bands compared to W3.

\begin{itemize}
    \item \textbf{Deep Tier 1}: An idealized deep tier with no magnitude error and infinite exposure time, corresponding to an arbitrarily deep limiting magnitude. In principle, such a survey would provide access to a large amount of non-linear information in the power spectrum, since a survey with infinite depth would detect arbitrarily faint objects and high-z galaxies, which are increasingly clustered. It is also the perfect, albeit unrealistic, deep tier for photo-z calibration. Although this tier assumes idealized parameter values for $\sigma_\mathrm{mag}$, $t_\mathrm{exp}$, and mag depth, its survey area remains $20 \deg^2$ and is therefore cosmic-variance limited. 
    
    \item \textbf{Deep Tier 2 \& 4}: The deep tiers D2 and D4 are $4\times$ longer than W2 over $20\deg^2$, being $+1$ and $+2$ deeper than W2, respectively.
    
    \item \textbf{Deep Tier 3}: This tier enables relatively easier cosmological constraints and photo-z calibration, as it has a shallower depth ($1$ mag shallower than D4) and covers a wider area ($20 \deg^2$ larger than D124). The increased area helps reduce cosmic variance in the power spectrum by including more information from linear scales, while the shallower depth excludes the faintest objects detected in D4. This deep tier is better constrained than the D124 deep tiers.
\end{itemize}

\begin{figure}[h!]
    \centering
    \includegraphics[width=1\linewidth]{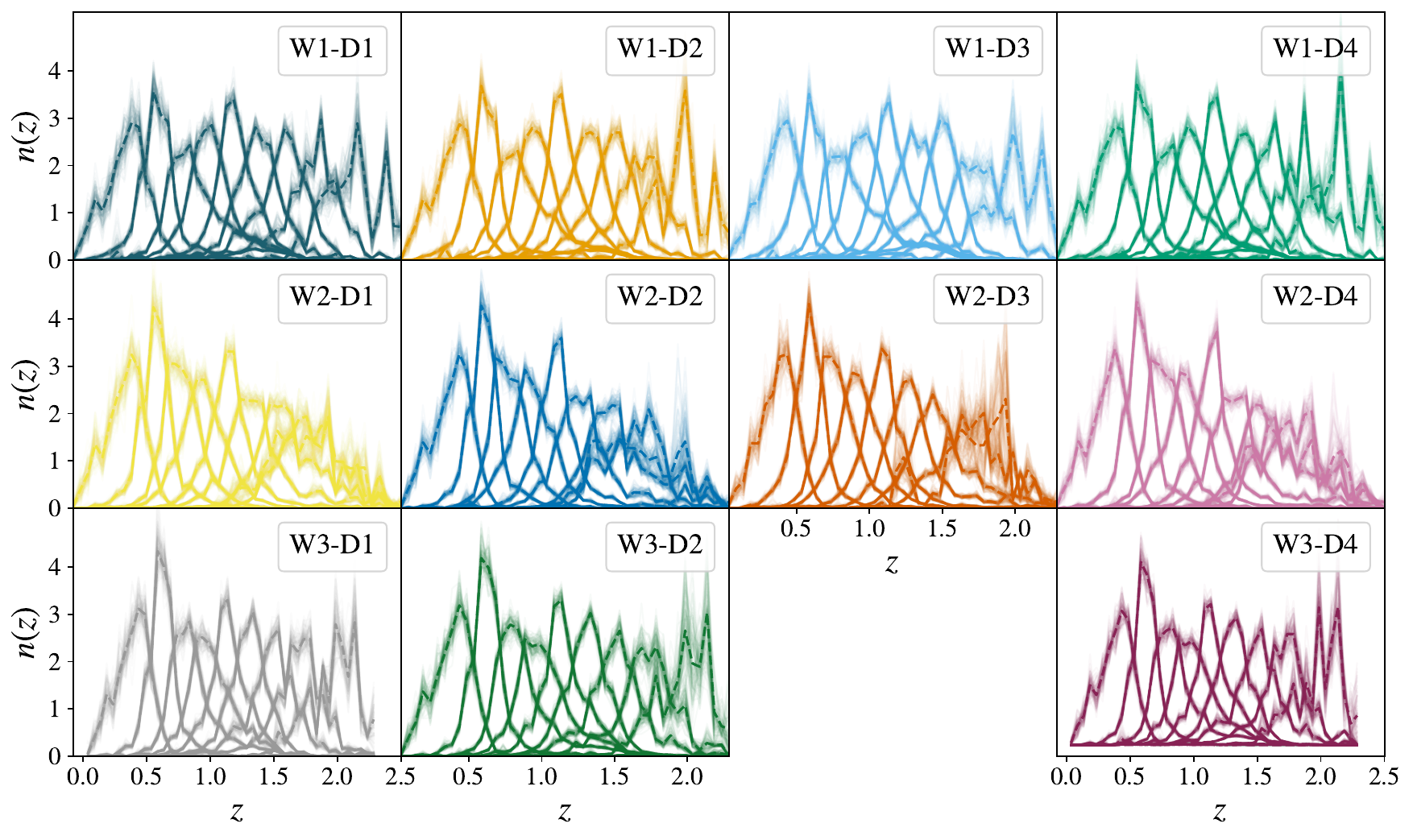}
    \caption{Estimated redshift distributions for different wide- and deep-tier specifications. Lighter lines display a small illustrative subsample of 100 (out of 1 million) realizations, with their spread reflecting the uncertainty in their exact shape, whereas the darker dashed lines show the mean redshift distribution computed over all 1 million realizations. Each row pairs a fixed wide tier (W1-W3) with a deep tier (D1-D4), except for the combination W3-D3.}
    \label{fig:roman_scenarios}
\end{figure}

\section{Modeling the shape uncertainty of redshift distributions}\label{sec:modeling_pca}

\subsection{Constructing the basis functions}\label{sec:unweighted_pca}

Let $\mathbf{n}_r$ denote a single realization from the ensemble of $N_{\mathrm{sim}} = 1{,}000{,}000$ redshift distributions generated with \texttt{Cardinal} (see Figure~\ref{fig:roman_scenarios}). Each realization is defined over $N_t = 9$ tomographic bins (indexed by $b$) and $N_z = 46$ redshift points (indexed by $k$) per tomographic bin. To represent $\mathbf{n}_r$ as a single vector, we concatenate the values from all tomographic bins, yielding a vector of length $N_{\mathrm{data}} \equiv N_t N_z = 414$. Stacking all $N_{\mathrm{sim}}$ realizations column-wise produces the matrix $\mathbf{n}$ of dimension $N_{\mathrm{data}} \times N_{\mathrm{sim}}$. Thus, each column of $\mathbf{n}$ corresponds to one realization $\mathbf{n}_r$, and each element $n_{r b k}$ is uniquely specified by the realization index $r$, tomographic-bin index $b$, and redshift index $k$:
\begin{align}\label{eq:binnz}
    \mathbf{n}_r&\equiv
    \{
    n_{rbk}| 
    r \in 1,\!\cdot\!\cdot\!\cdot, N_\mathrm{sim}|
    b \in 1,\!\cdot\!\cdot\!\cdot, N_t| 
    k \in 1,\!\cdot\!\cdot\!\cdot,N_z
    \}.
\end{align}

For a fixed tomographic bin index $b$ and redshift point $k$, we have 1 million values of $n_{rbk}$, whose variance reflects the uncertainty in their determination. We apply the principal component method to linearly transform the data points $n_{rbk}$ into a new coordinate system, such that the directions (principal components, or modes) correspond to the axes of maximal variance in the data.

\begin{figure}[h]
    \centering
    \includegraphics[width=\linewidth]{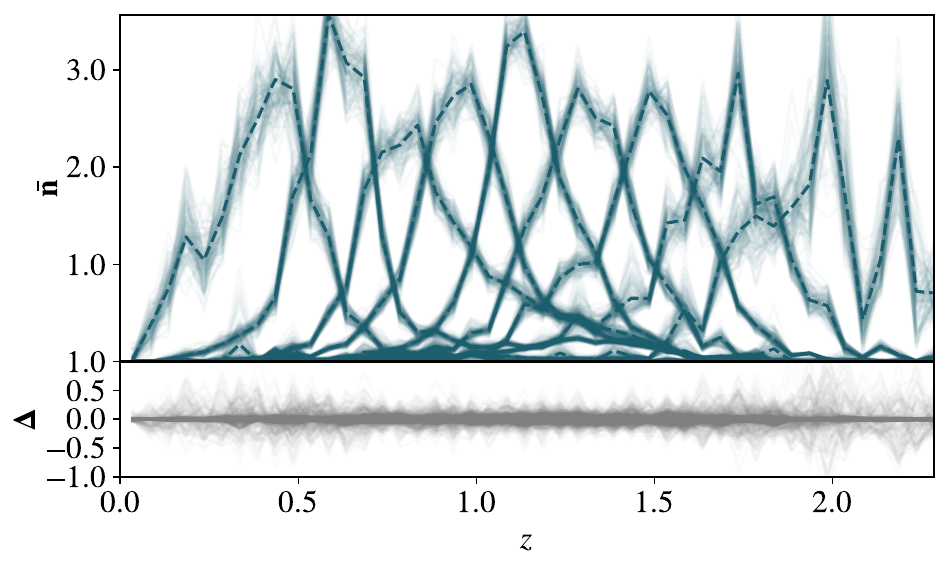}
    \caption{Quantities derived from the Roman scenario W1-D1. \textit{Top}: mean redshift distribution (dark dashed lines) obtained from the ensemble of 1 million realizations, along with 100 individual realizations (faint lines). \textit{Bottom}: difference between the mean redshift distribution and the individual realizations per tomographic bin.}
    \label{fig:redshift_delta_sc1bd4_v2}
\end{figure}

The first step is to center the data around the mean redshift distribution $\bar{\mathbf{n}}$ by computing the difference matrix $\mathbf{\Delta}$, defined as the difference between each realization and the mean
\begin{align}\label{eq:delta}
\bar{\mathbf{n}}&\equiv\langle\mathbf{n}\rangle_\mathrm{sim}=\frac{1}{N_\mathrm{sim}}\sum_{r=1}^{N_\mathrm{sim}} \mathbf{n}_r,\nonumber\\
\mathbf{\Delta} &\equiv \mathbf{n} - \bar{\mathbf{n}} =
\left[
\mathbf{n}_{r=1} - \bar{\mathbf{n}} \;\big|\; \mathbf{n}_2 - \bar{\mathbf{n}} \;\big|\; \cdots \;\big|\; \mathbf{n}_{N_\mathrm{sim}} - \bar{\mathbf{n}}
\right]_{N_\mathrm{data} \times N_\mathrm{sim}}.
\end{align}
The length of $\bar{\mathbf{n}}$ is $N_\mathrm{data}$, while $\mathbf{\Delta}$ has the same dimensions as $\mathbf{n}$, namely $N_\mathrm{data} \times N_\mathrm{sim}$. Figure~\ref{fig:redshift_delta_sc1bd4_v2} is an example of these two quantities for the Roman scenario W1-D1. The top panel is the mean redshift distribution (the same mean as shown in Figure~\ref{fig:roman_scenarios}), and the bottom panel are the first 100 columns of the difference matrix arranged by the $N_t$ tomographic bins.

The principal components are computed from the singular value decomposition (SVD) of the difference matrix
\begin{align}\label{eq:delta_svd}
    \mathbf{\Delta} = \mathbf{D} \mathbf{\Sigma} \tilde{\mathbf{U}}^T.
\end{align}
The factorization in Equation \eqref{eq:delta_svd} is illustrated in Figure~\ref{fig:redshift_delta_sc1bd4_v2_svd} (cf. Fig.~5 of the baryon PCA used by \cite{baryon_pca_wl_2}). In this decomposition, the \textit{decoding} matrix $\mathbf{D}$ and the matrix $\tilde{\mathbf{U}}^T$ are unitary and have dimensions $N_\mathrm{data} \times N_\mathrm{data} = 414 \times 414$ and $N_\mathrm{sim} \times N_\mathrm{sim} = 10^6 \times 10^6$, respectively. The matrix $\mathbf{\Sigma}$ has dimensions $N_\mathrm{data} \times N_\mathrm{sim} = 414 \times 10^6$, whose upper $414 \times 414$ block contains the positive real singular values $\sigma_1, \sigma_2, \ldots, \sigma_{414}$ in decreasing order, and the remaining entries are zero, as indicated by the dashed square. The columns of $\mathbf{D}$ correspond to the PCs (or eigenvectors) of the covariance matrix 
\begin{align}
    \mathbf{C}_\mathbf{\Delta} \equiv \mathbf{\Delta} \mathbf{\Delta}^T,
\end{align}
with the eigenvalues given by the diagonal elements of $\mathbf{\Sigma} \mathbf{\Sigma}^T$. Interestingly, the SVD of $\mathbf{\Delta}$ is equivalent to the eigendecomposition of ``${C}_n$'' used in \cite{pca_method}. To keep the notation used here as close as possible to \cite{pca_method}, we denote PCs modes by $\mathbf{U}_i$. To represent any redshift distribution in the basis of the principal components, we use the first $M(\leq N_\mathrm{data})$ columns $\mathbf{U}_i$ of the matrix $\mathbf{D}$ 

\begin{figure}
\centering
\begin{tikzpicture}[
    font=\sffamily,
    box/.style={draw, minimum height=1cm, minimum width=3cm, fill=gray!20},
    widebox/.style={draw, minimum height=4cm, minimum width=10cm, fill=gray!20},
    darkbox/.style={draw, minimum height=.96cm, minimum width=3cm, fill=gray!60},
    darkbox2/.style={draw, minimum height=.9cm, minimum width=5cm, fill=gray!60},
    label/.style={font=\small},
    tlabel/.style={font=\bfseries\Large}
  ]

\node[tlabel] at (0, 2.1) {$\mathbf{\Delta}$};
\node[tlabel] at (2.04, 0.0) {$=$};
\node[tlabel] at (4.2, 2.1) {$\mathbf{D}$};
\node[tlabel] at (7.6, 2.1) {$\mathbf{\Sigma}$};
\node[tlabel] at (11.4, 2.1) {$\tilde{\mathbf{U}}^T$};

\node[box] (delta) at (0,0) {};
\node[widebox, anchor=west, minimum width=3.5cm, minimum height=3cm] (uRest) at (2.55,0) {};
\node[darkbox,rotate=90] (pcblock) at (4.302,0) {};
\node[darkbox2, minimum width=3.5cm] (pcblock) at (4.310,0) {};
\node[box, fill=white, minimum height=1cm, minimum width=3cm, right=0.5cm of uRest] (sigma) {};
\node[box, dotted, fill=none, minimum height=1.3cm, line width=1pt, minimum width=1.34cm, right=2.3cm of uRest] (test) {};
\node[box, dotted, minimum height=3cm, minimum width=3.2cm, right=.5cm of sigma] (vt) {};

\node[text=white] at (pcblock) {
  $~\mathbf{U}^b_1 \displaystyle\cdots 
  \begin{array}{c}
    ~~\vdots \\[-1pt]
    \mathbf{U}_i^b(z) \\[-1pt]
    ~~\vdots
  \end{array}
  \displaystyle\cdots \mathbf{U}_{414}^b~$
};
\node[text=white] at (pcblock) at (4.31,1.2) {${\mathbf{U}^1_i}(z)$};
\node[text=white] at (pcblock) at (4.31,-1.2) {$\mathbf{U}^9_i(z)$};
\node[text=black, anchor=north west] at (pcblock) at (uRest.north west) {$46~\{$};
\node[text=black, anchor=south west] at (pcblock) at (uRest.south west) {$46~\{$};

\draw[line width=0.6pt, shorten <=18pt, shorten >=25pt] (sigma.north west) -- (8,-.5);
\node[label, anchor=north west] at (sigma.north west) {$\sigma_1$};
\node[label, anchor=south east] at (8,-.5) {$\sigma_{414}$};

\node[below=0.4cm of delta] {$\substack{\displaystyle\scriptstyle N_\text{data} \times N_\text{sim} ~\\ \displaystyle (414\times1\text{M})}$};
\node[below=0.4cm of uRest] {$\substack{\displaystyle\scriptstyle N_\text{data} \times N_\text{data} \\ \displaystyle (414\times414)}$};
\node[below=0.4cm of sigma] {$\substack{\displaystyle\scriptstyle N_\text{data} \times N_\text{sim} ~\\ \displaystyle (414\times1\text{M})}$};
\node[below=0.4cm of vt] {$\substack{\displaystyle\scriptstyle N_\text{sim} \times N_\text{sim} \\ \displaystyle (1\text{M}\times1\text{M})}$};

\end{tikzpicture}
\caption{Singular value decomposition of the difference matrix $\mathbf{\Delta}$ for the $N_\text{sim}=10^6$ realizations for the redshift distributions. The unitary matrix $\mathbf{D}$ holds the left singular vectors and $\tilde{\mathbf{U}}^T$ contains the right singular vectors, while $\mathbf{\Sigma}$ possesses the singular values; the upper \(414\times414\) block contains the positive real singular values \(\sigma_1,\ldots,\sigma_{414}\) in decreasing order, and the dashed square indicates that all remaining entries are zero.}
\label{fig:redshift_delta_sc1bd4_v2_svd}
\end{figure}
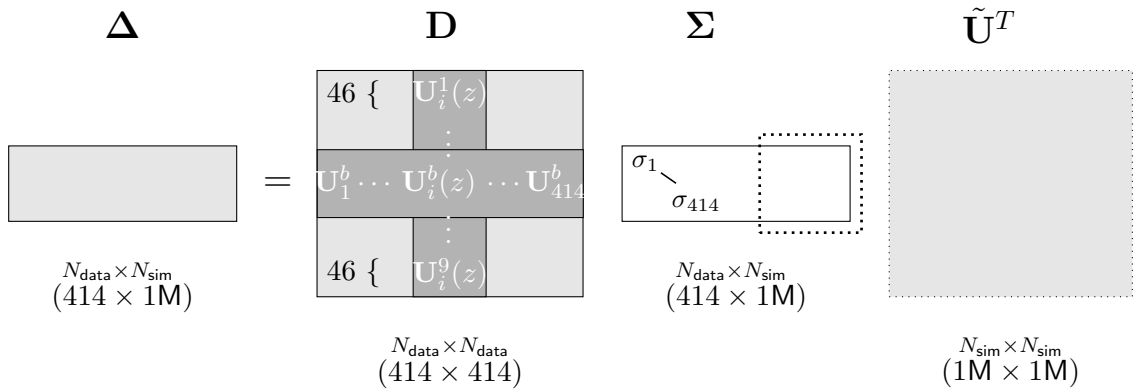
\begin{align}\label{eq:n_in_e_base}
    \hat{\mathbf{n}} = \bar{\mathbf{n}} + \sum_{i=1}^{M}  u_i \mathbf{U}_i,
\end{align}
where $u_i$ are the amplitudes (or projection coefficients) of the PCs. In Fig.~\ref{fig:redshift_delta_sc1bd4_v2_svd} the \textit{i}th PC in the \textit{b}th tomographic bin is denoted by $\mathbf{U}_i^b(z)$ where the indexes range are $i=1,\cdots,N_\mathrm{data}$ and $b=1,\cdots,N_t$. In Fig.~\ref{fig:pc_example}, we present a two-dimensional example of the PCA method applied to the ensemble of realizations derived from Equation \eqref{eq:binnz}:
\begin{figure}[h]
    \centering
    \includegraphics[width=\linewidth]{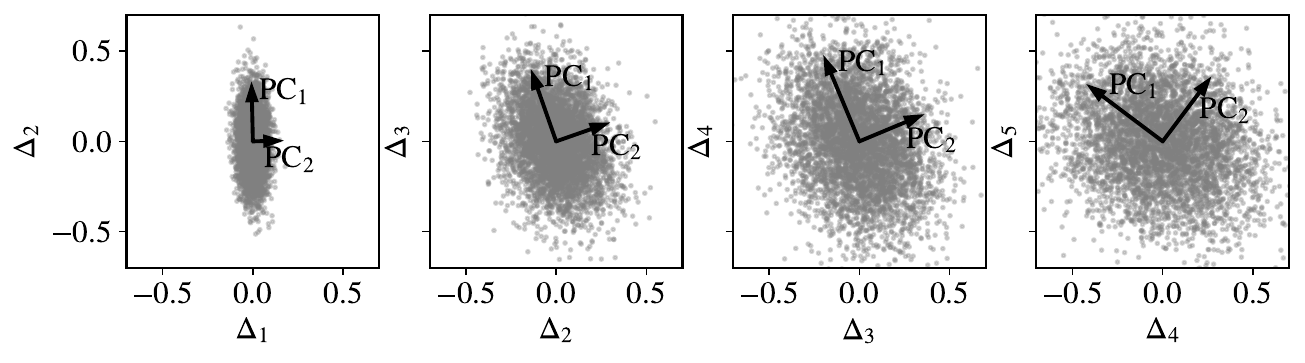}
    \caption{Two-dimensional projection of the ensemble of the SVD application to the redshift distributions of Roman scenario W1-D1. Gray points represent draws from a Gaussian distribution with mean zero and a covariance matrix. The correlations between dimensions are captured by the covariance of the difference vectors $\Delta_i$. The arrows indicate the first ($\mathbf{PC}_1$) and second ($\mathbf{PC}_2$) principal components.}
    \label{fig:pc_example}
\end{figure}

The PCA approach presented so far constructs basis functions directly from the SVD of the ensemble of realizations around the mean redshift distribution. However, it does not account for how each realization impacts the cosmic shear power spectra. Consequently, the inferred cosmological parameters are only weakly sensitive to the redshift-distribution uncertainty captured by these unweighted principal components. An optimized, signal-aware PCA incorporating weights that encode the response of shear observables to fluctuations in $\mathbf{n}$ is introduced in the next section, following \cite{pca_method}.

\subsection{Optimizing the basis function with respect to differences in the correlation function}\label{sec:weighted_pca} 

Variation of the cosmic shear power spectra with respect to the ensemble of redshift distribution can be used to optimize the PC basis presented in Section \ref{sec:unweighted_pca}. We follow the prescription of \cite{pca_method} which we briefly summarize here for completeness: 

First, get the mean of the ensemble of the redsfhit distributions $\bar{\mathbf{n}}$. Next, compute the Jacobian matrix by taking derivatives of $\xi_\pm$ with respect to the redsfhit distributions rescaled by the mean. Finally, determine the cosmic shear covariance matrix. These steps allow to build the following Fisher matrix:

\begin{align}\label{eq:fisher_matrix}
    \mathbf{F} \equiv \left[ \frac{\partial \xi_\pm}{\partial \mathbf{n}} \right]_{\bar{\mathbf{n}}}^{T} \mathbf{C}_{\xi_\pm}^{-1} \left[ \frac{\partial \xi_\pm}{\partial \mathbf{n}} \right]_{\bar{\mathbf{n}}}.
\end{align}

To compute the cosmic shear covariance matrix, $\mathbf{C}_{\xi_\pm}$, we employ \texttt{CosmoCov}~\cite{cosmocov_1,cosmocov_2} to evaluate the Gaussian\footnote{We adopt the Gaussian covariance as a controlled approximation for this validation study of the redshift-distribution PCA implementation within \texttt{CoCoA}. A full treatment including non-Gaussian and super-sample covariance terms would substantially increase the computational cost for the nine tomographic bins considered in this work, but should be included in future survey analyses.} covariance matrix for the $3\times2$ point-correlation function. We used 15 angular bins between $\theta_\mathrm{min} = 2.5~\mathrm{arcmin}$, and $\theta_\mathrm{max} = 250~\mathrm{arcmin}$. The \texttt{CoCoA} framework is then used to derive the masked covariance matrix for the cosmic shear component, $\mathbf{C}_{\xi_\pm}$. For the Jacobian matrix $\mathbf{J}$ defined as:  
\begin{align}\label{eq:jacobian_matrix}
    \mathbf{J} \equiv \frac{\partial \xi_\pm}{\partial \mathbf{n}},
\end{align}
we employ the five-point stencil forward finite difference method:  
\begin{align}
    f'(x) \approx \frac{1}{12\epsilon} \Big[-25 f(x) + 48 f(x+\epsilon) - 36 f(x+2\epsilon) + 16 f(x+3\epsilon) - 3 f(x+4\epsilon) \Big], \nonumber
\end{align}
where $f \rightarrow \xi_\pm(\theta;x)$ and $x \rightarrow {n}_{rbk} = \mathbf{n}_r(z_k)$ at the $b$th tomographic bin. The one-sided forward derivative is chosen to prevent negative shifts in $x$, since it represents redshift distributions. 

The weights to optimize the PC basis is derived from the Fisher matrix in Equation \eqref{eq:fisher_matrix} by performing additional matrix transformation, which describes the mapping from the original space of redshift distributions to the PC space. First, compute the eigenvalues $\mathbf{\Lambda}_n$ and the eigenvectors $\mathbf{V}n$ of the sample covariance matrix $\mathbf{C}{\mathbf{\Delta}} = \mathbf{\Delta}\mathbf{\Delta}^T$. Then, build the transformed Fisher matrix $\mathbf{G}\equiv\mathbf{\Lambda}_n^{1/2}\mathbf{V}_n^T\mathbf{F}\mathbf{V}_n\mathbf{\Lambda}_n^{1/2}$ and find its eigenvalues $\mathbf{\Lambda}_G$ and eigenvectors $\mathbf{V}_G$. With these ingredients, one can construct a projection matrix $\mathbf{P}_M$ that is 1 for the first $M$ eigenmodes of $\mathbf{G}$ (ordered by decreasing eigenvalue) and 0 otherwise, where $M$ defines the number of retained, most informative modes. With these matrices, we can then derive the \textit{encoding} $\mathbf{E}\equiv \mathbf{P}_M\mathbf{V}_G^T\mathbf{\Lambda}_n^{-1/2}\mathbf{V}_n^T$ and the \textit{decoding} $\mathbf{D}\equiv\mathbf{V}_n^T\mathbf{\Lambda}_n^{1/2}\mathbf{V}_G\mathbf{P}_M$ operators used to project $\mathrm{n}(z)$ onto the leading PC and to reconstruct the redshift distribution, respectively. We defer further details of this formalism to \cite{pca_method} and \cite{des_y6_pca_redshift_calib_source,des_y6_pca_redshift_calib_lens} for applications in DES Y6.

\begin{figure}[h]
    \centering
    \includegraphics[width=\linewidth]{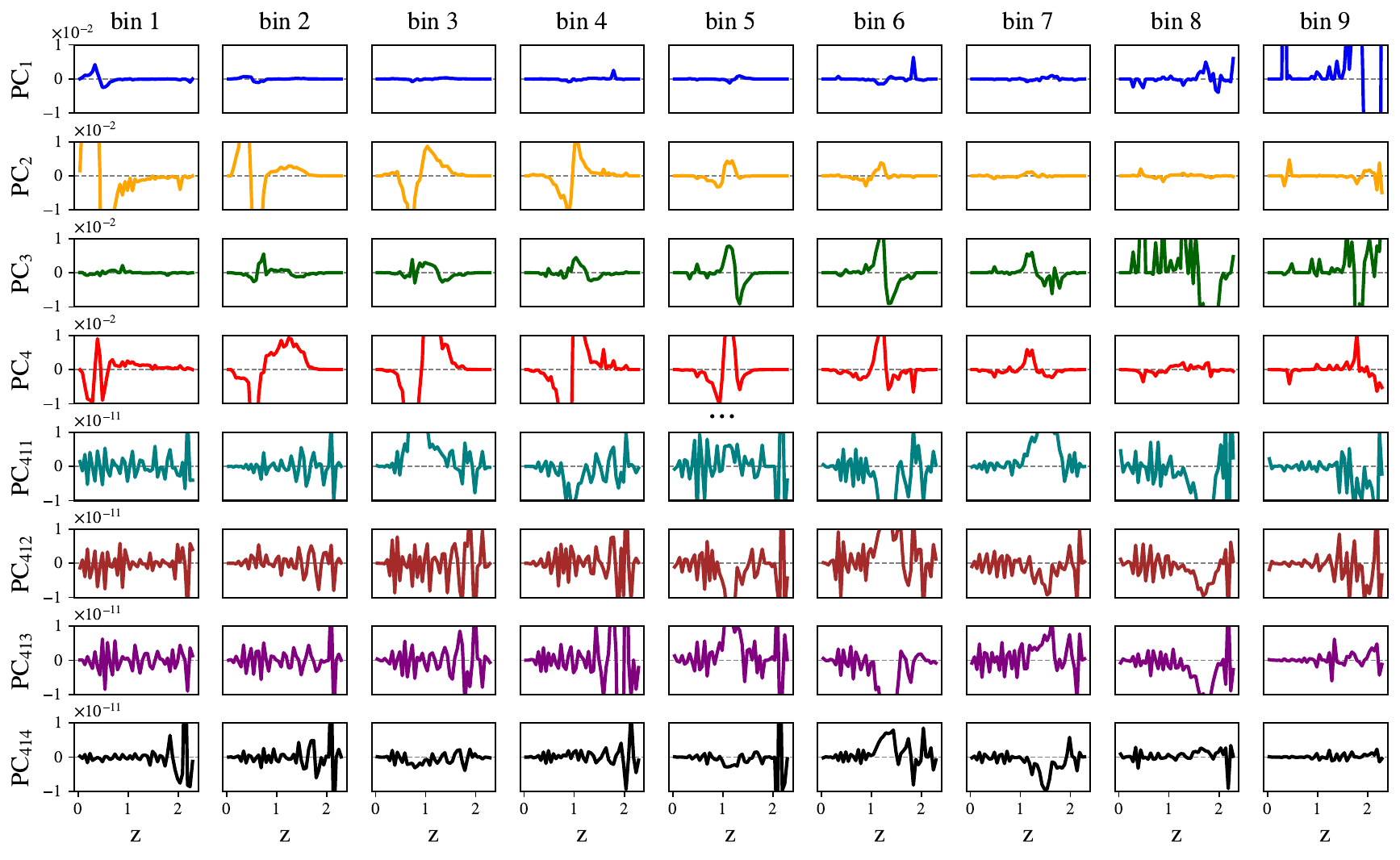}
    \caption{The first four and the last four PC modes, as function of the redshift, across the nine tomographic redshift bins for the Roman scenario W1-D1.}
    \label{fig:PCs_DRMD1_weighted}
\end{figure}

Fig.~\ref{fig:PCs_DRMD1_weighted} presents the first four principal components ($\mathrm{PC}_1,\ldots,\mathrm{PC}_4$) and the last four $(\mathrm{PC}_{411},\ldots,\mathrm{PC}_{414})$ derived from the Roman W1-D1 scenario, separated according to their contributions across the nine tomographic bins. The DES \cite{des_y6_pca_redshift_calib_source} also provide their principal components in Fig.~E2. The impact of each principal component on the redshift distribution should be evaluated using its individual amplitude $u_i$, since, according to Equation \eqref{eq:n_in_e_base}, the contribution scales with the product $u_i\,\mathbf{U}_i$.

The distribution of the projected amplitudes $ u_i$s for the ensemble of 1 million redshift distribution can be obtained by applying the encoding matrix (see \cite{pca_method})
\begin{align}\label{eq:projection}
    \mathbf{u}=\mathbf{E}\mathbf{\Delta},
\end{align}
or by projecting the difference matrix $\hat{\mathbf{\Delta}}=\hat{\mathbf{n}}-\bar{\mathbf{n}}$ onto the PCs as per Equation \eqref{eq:n_in_e_base} and summing over every redshift point and tomographic bin
\begin{align}
    u_i = \langle \hat{\mathbf{\Delta}}, \mathbf{U}_i \rangle = \sum_{b=1}^{N_t} \sum_{k=1}^{N_z} \hat{\mathbf{\Delta}}_{bk} \cdot \mathbf{U}_{bki}.
\end{align}

\begin{figure}
    \centering
    \includegraphics[width=0.6\linewidth]{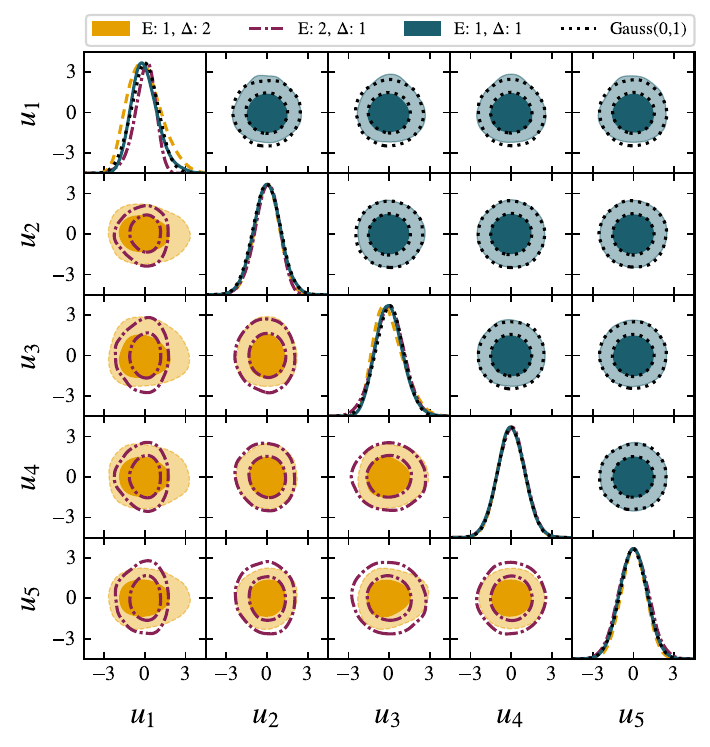}
    \caption{Prior probability distributions of the first five projected amplitudes, $u_i$, along with Gaussian fits shown as dotted black lines. \textit{Upper triangle:} both the encoding matrix and the difference matrix are derived from the same Roman scenario, W1-D1. \textit{Lower triangle:} the yellow shaded region indicates that the encoding matrix is derived from W1-D1 and the difference matrix from W2-D2, while the purple dash-dotted line represents the reverse case, where the encoding matrix is derived from W2-D2 and the difference matrix from W1-D1.}
    \label{fig:alphas_distributions_double}
\end{figure}

The prior probability distributions of the first five projected amplitudes, $u_i$, for the Roman scenario W1-D1 are shown in the upper-triangle plot of Figure~\ref{fig:alphas_distributions_double} (dark green region). The label ``E: 1, $\Delta$: 1'' indicates that both the encoding matrix $\mathbf{E}$ and the ensemble of centered realizations $\mathbf{\Delta}$ are derived from the same Roman scenario W1-D1 (see Equation \eqref{eq:projection}). As expected, the amplitudes are uncorrelated and centered around zero, making the $u_i$ well approximated by a Gaussian distribution with mean zero and unit variance, represented by the dotted black lines in Figure~\ref{fig:alphas_distributions_double}. In Section~\ref{sec:results}, we use this prior information in our MCMC analysis. In principle, however, $\mathbf{E}$ and $\mathbf{\Delta}$ can be drawn from different Roman scenarios, shown in the lower triangle plot of Fig.~\ref{fig:alphas_distributions_double}. For instance, the configuration ``E: 2, $\Delta$: 1'' corresponds to principal components derived from a noisier Roman scenario W2-D2 applied to a synthetic cosmic shear data vector based on the idealized W1-D1 scenario (see Table~\ref{tab:roman_scenarios}), and vice versa for ``E: 1, $\Delta$: 2''. Mixing the Roman scenarios slightly reduces the Gaussianity of the amplitude distributions, a case we examine in more detail in Section~\ref{sec:results}.

{\begin{figure}[h]
\centering
\begin{tikzpicture}[node distance=12mm and 10mm, every node/.style={transform shape}]
\node[box, text width=3.0cm] (real) {\textbf{Realizations \\}$\mathbf{n}^r(z)$
};
\node[smallbox, below=of real] (mean) {\textbf{Mean }$\bar{\mathbf{n}}(z)$ $\bar{\mathbf{n}}(z)=\frac{1}{N_\mathrm{sim}}\sum_r \mathbf{n}_r(z)$};
\node[smallbox, below=of mean] (delta) {\textbf{Difference matrix }$\mathbf{\Delta}(z)$\\$\mathbf{\Delta}(z)=\mathbf{n}(z)-\bar{\mathbf{n}}(z)$};
\node[smallbox, below=of delta] (cov) {\textbf{Covariance }\\$\mathbf{C}_\mathbf{\Delta}=\mathbf{\Delta}\,\mathbf{\Delta}^{\!T}=\mathbf{V}\mathbf{\Lambda}\mathbf{V}^T$};
\node[box, text width=4.0cm,  right=of real] (cocoa_cosmocov) {\textbf{CosmoCov \& CoCoA}\\ Compute covariance $\mathbf{C}$\\Compute observable $\xi_{\pm}(\theta)$\vspace{2pt}};
\node[smallbox, text width=3.0cm,  below=of cocoa_cosmocov] (jacobian) {\textbf{Jacobian}\\$\mathbf{J}\equiv\partial\xi_{\pm}/\partial{\mathbf{n}}(z)$};
\node[smallbox, text width=3.0cm,  below=of jacobian] (fisher) { \textbf{Fisher}\\$\mathbf{F}=\mathbf{J}^{T}\,\mathbf{C}_{\xi_\pm}^{-1}\,\mathbf{J}$};
\node[box, text width=4.0cm, right=of cocoa_cosmocov] (model) {\textbf{Compressed redshift distribution} $n(z)=\bar{n}(z)+\sum_k u_k\,\mathrm{PC}_k(z)$};
\node[smallbox, text width=4.0cm, below=of model] (mcmc) {\textbf{Analysis with CoCoA}\\MCMC or Evaluation};
\node[smallbox, text width=4.0cm, below=of fisher] (wpca) {\textbf{Weighted PCA}\\Diagonalize weighted matrix\\$\mathbf{G}=\mathbf{\Lambda}^{1/2}\mathbf{V}^{T}\,\mathbf{F}\,\mathbf{V}\mathbf{\Lambda}^{1/2}$\\Eigenvectors $\rightarrow$ PC$_k(z)$};
\node[smallbox, below=of mcmc, text width=4.5cm, inner sep=4pt] (contour) {
    \textbf{Posteriors}\\[2pt]
    \includegraphics[width=4.2cm]{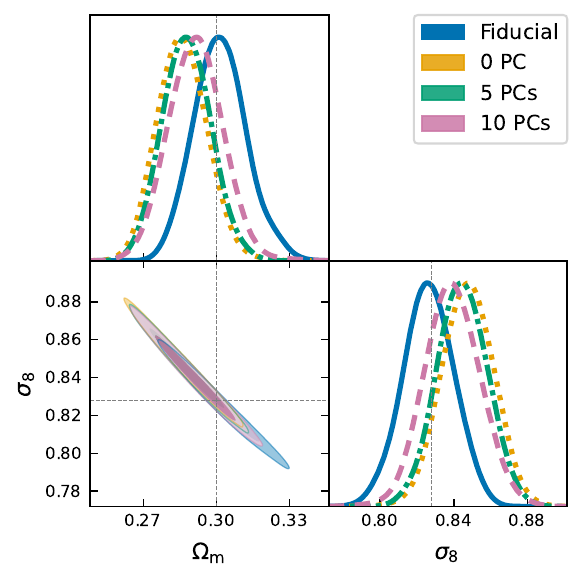}
};
\draw[arrow] (real.south) -- (mean.north);
\draw[arrow] (mean.south) -- (delta.north) ;
\draw[arrow] (real.west) -- ++(-5mm,0) |- (delta.west);
\draw[arrow] (delta.south) -- (cov.north);
\draw[arrow] ([xshift=19.9mm]cov.south) |- (wpca.west);
\draw[arrow] (mean.east) -- ++(3.5mm,0) |- (cocoa_cosmocov.west) ;
\draw[arrow] (model.south) -- (mcmc.north) ;
\draw[arrow] (mcmc.south) -- (contour.north);
\draw[arrow] (wpca.east) -- ++(5mm,0) |- (model.west) node[midway, font=\footnotesize, text=muted, rotate=90, yshift=2mm, xshift=-40mm] {select leading PCs} ;
\draw[arrow] (cocoa_cosmocov.south) -- (jacobian.north);
\draw[arrow] (jacobian.south) -- (fisher.north);
\draw[arrow] (fisher.south) -- (wpca.north);
\end{tikzpicture}
\caption{Pipeline workflow for the $n(z)$ PCA method implemented in \texttt{CoCoA}. 
    }
    \label{fig:pca-pipeline}
\end{figure}}

Having established all the necessary components, we are now equipped to perform a likelihood analysis using the new PCA framework. The main steps of the (un)weighted PCA pipeline are summarized and illustrated in Figure~\ref{fig:pca-pipeline}:  
(\textit{i}) Using \texttt{Cardinal}, we generate one million realizations of the redshift distributions.  
(\textit{ii}) Compute the covariance matrix $\mathbf{C}_\mathbf{\Delta}$ of the samples, centered around the mean distribution $\bar{\mathbf{n}}$. The square matrix $\mathbf{C}_\mathbf{\Delta}$ is then decomposed into its eigenvectors $\mathbf{V}$ and eigenvalues $\mathbf{\Lambda}$. Stopping here yields the unweighted PCA method, where the PCs are given by $\mathbf{V}$ (or equivalently, by the $\mathbf{D}$ matrix derived from the SVD of $\mathbf{\Delta}$).  
(\textit{iii}) With \texttt{CosmoCov} and \texttt{CoCoA}, compute the covariance of the observable $\xi_\pm$ and the Jacobian matrix $\mathbf{J} = \partial\xi_\pm / \partial\mathbf{n}$.  
(\textit{iv}) Combine $\mathbf{J}$ with the inverse of $\mathbf{C}_{\xi_\pm}$ to obtain the Fisher matrix $\mathbf{F}$.  
(\textit{v}) Derive the transformed Fisher matrix $\mathbf{G}$ (see \cite{pca_method}), from which we construct the ``encoding'' matrix $\mathbf{E}$ and the ``decoding'' matrix $\mathbf{D}$.  
(\textit{vi}) The columns of $\mathbf{D}$ represent the principal components (modes) that define the PCA model for the redshift distribution.  
(\textit{vii}) In the MCMC analysis, we sample over the PC amplitudes $u_i$ to assess their impact on cosmological parameter constraints (e.g., $\sigma_8$ versus $\Omega_m$). The pipeline shown in Fig.~\ref{fig:pca-pipeline} is implemented in \texttt{CoCoA} and available at \texttt{cocoa\_photoz}\footnote{\url{https://github.com/Roman-HLIS-Cosmology-PIT/cocoa_photoz}}.

\section{Analysis methodology}\label{sec:analysis_methodology}

The posterior distributions of the cosmological parameters are inferred through a Markov Chain Monte Carlo (MCMC) approach, with convergence evaluated using the Gelman-Rubin statistic, requiring $R - 1 < 0.015$. The Bayesian analysis is performed with \texttt{Cobaya}\footnote{\url{https://github.com/CobayaSampler/cobaya}}~\cite{cobaya}, while theoretical predictions are computed using \texttt{CAMB}\footnote{\url{https://github.com/cmbant/camb}}~\cite{camb}
 and \texttt{CosmoLike} within the \texttt{CoCoA} framework. The resulting MCMC chains are analyzed and visualized with \texttt{GetDist}\footnote{\url{https://github.com/cmbant/getdist}}~\cite{getdist}.

The synthetic cosmic shear power spectra is generated given a redshift distribution (e.g., any normalized $n(z)$ shown in Figure~\ref{fig:roman_scenarios}), and the fiducial cosmological parameters listed in Table~\ref{tab:fiducial_cosmology}. The parameters varied in the MCMC chains are the total matter density $\Omega_m$, baryon density $\Omega_b$, Hubble constant $H_0$, amplitude of primordial fluctuations $A_s$, and scalar spectral index $n_s$. The optical depth $\tau$ is held fixed, and the neutrino sector is set to the default \texttt{CAMB} configuration-one massive neutrino with $m_\nu = 0.06~\mathrm{eV}$, degenerate mass eigenstates, and an effective number of neutrinos $N_\mathrm{eff} = 3.044$. For the $\Lambda$CDM model, the Chevallier-Polarski-Linder (CPL) parameters are fixed to $w_0 = -1$ and $w_a = 0$. The Survey parameters are the total sky are, $\Omega_s$, the total shape noise of weak lensing measurements, $\sigma_\epsilon$, and the number density in each bin, $n_g$. In this work, we consider two models according to Table~\ref{tab:fiducial_cosmology}:  
(\textit{i}) $\Lambda$CDM+Shift - the standard analysis in which the amplitudes of the principal components are fixed to zero while the mean-shift parameters are varied; and  
(\textit{ii}) $\Lambda$CDM+PCA - the case where the mean-shift parameters are fixed in the chain and the amplitudes of the principal components are varied. Note that the index \textit{i} associated with $\Delta_z^i$ refers to the tomographic bins, whereas when associated with $u_i$, it denotes the \textit{i}th principal component amplitude. Each PC $\mathbf{U}_i$ has a single amplitude $u_i$ shared across all nine tomographic bins; however, the product $u_i \mathbf{U}_i$, as defined in Equation \eqref{eq:n_in_e_base}, differs between bins because the intrinsic shape of $\mathbf{U}_i$ varies with tomographic bin.

\begin{table}[h!]
\centering
\begin{tabular}{l@{\hspace{2.5em}}l@{\hspace{1.2em}}l}
\toprule
\textbf{Parameter} & \textbf{Fiducial} & \textbf{Prior} \\
\midrule
\multicolumn{3}{l}{\textbf{Cosmology}} \\
$\Omega_m$ & 0.3 & flat (0.1, 0.6) \\
$\Omega_b$ & 0.04 & flat (0.04, 0.055) \\
$H_0$ & 67.32 & flat (60, 76) \\
$10^9A_s$  & 2.1   & flat (1.1,2.8)\\
$n_s$ & 0.966 & flat (0.87, 1.07) \\
$\sum m_\nu$ & 0.06 & fixed \\
$\tau$ &0.0697 & fixed\\
$w_0$ & -1 & fixed\\
$w_a$ & 0 & fixed\\
\midrule
\multicolumn{3}{l}{\textbf{Survey}} \\
$\Omega_s$ & 2000 deg$^2$ & fixed \\
$\sigma_\epsilon$ & 0.39 & fixed \\
$n_g$ & $40~\text{gal}/\text{arcmin}^2$ & fixed \\
\midrule
\multicolumn{3}{l}{\textbf{Shear calibration}} \\
$m_i, i\in[1,9]$ & 0 & Gauss $(0.0, 0.005)$ \\
\midrule
\multicolumn{3}{l}{\textbf{Intrinsic Alignment}} \\
$a_\mathrm{IA}$ & 0 & flat $(-5, 5)$ \\
$\eta$ & 0 & flat $(-5, 5)$ \\
\midrule
\multicolumn{3}{l}{\textbf{Source photo-z}} \\
$\Delta_z^i, i\in[1,9]$ & 0 & Gauss (a): $(0,0.003)$\\
$\Delta_z^i, i\in[1,9]$ & 0 & Gauss (b): $(0,0.01)$\\
\midrule
\multicolumn{3}{l}{\textbf{PCA}} \\
$\Delta_z^i, i\in[1,9]$ & 0 & fixed\\
$u_i, i\in[1,M]$ & 0 & Gauss $(0,1)$ \\
\bottomrule
\end{tabular}
\caption{Fiducial values and priors for cosmological, survey, and systematic parameters. The Cosmology, Survey, and Shear Calibration blocks define the baseline $\Lambda$CDM model, the properties of our adopted fiducial Roman survey, and the shear calibration nuisance parameters, respectively. The Source photo-z and PCA blocks represent a modification of this baseline; when the former are varied in a MCMC, the latter are fixed to the fiducial values, and vice-versa.}
\label{tab:fiducial_cosmology}
\end{table} 

We use as the synthetic data vector the two tomographic cosmic shear correlation functions $\xi_\pm^{ij}(\theta)$, where the $\pm$ denotes the sum and difference of the tangential- and and cross-components of the projected galaxy shape distortion due to weak gravitational lensing. We compute the cosmic shear in 15 angular bins, uniformly spaced in logarithmic scale covering the range $\theta_\text{min} = 2.5~\mathrm{arcmin}$ to $\theta_\text{max} = 250~\mathrm{arcmin}$. 

The shear signal is a combination of the gravitational shear (hereafter $\kappa$) and intrinsic alignments (denoted by I), and these are decomposed into the leading- and higher-order weak lensing effects denoted by the E- and B-mode components, respectively. The connection between the correlation function in Real space with the power spectra in Fourier space is as follows \cite{des_y3_cosmology_from_clustering_and_wl,des_y3_cosmic_shear_robustness_to_calib}:
\begin{align}
    \xi_\pm^{ij}(\theta) = \sum\frac{2\ell+1}{2\pi\ell^2(\ell+1)^2} \left[G^+_{\ell,2}(\cos\theta)\pm G^-_{\ell,2}(\cos\theta)\right]\left[C_{EE}^{ij}(\ell) \pm C_{I_B I_B}^{ij}(\ell)\right],
\end{align}
where the two auxiliary functions $G^\pm_{\ell,m=2}(x)$ are obtained from the Legendre polynomials \cite{wl_celestial,des_y3_multi_probe}. Each E/B-mode term is  given by \cite{des_y3_multi_probe}:
\begin{align}\label{eq:CEEell}
    C_{EE}^{ij}(\ell) &= C_{\kappa\kappa}^{ij}(\ell) + C_{\kappa I_E}^{ij}(\ell) + C_{\kappa I_E}^{ji}(\ell) + C_{I_E I_E}^{ij}(\ell),\\
    C_{BB}^{ij}(\ell) &= C_{I_BI_B}^{ij}(\ell).
\end{align}
The gravitational shear, represented by the first term in Equation \eqref{eq:CEEell}, can be expressed in the Limber approximation as in \cite{limber_approx,des_y1_multiprobe,cov_modeling_des_y3}:
\begin{align}\label{eq:shear_spectra}
    C_{\kappa\kappa}^{ij}(\ell) = \int_0^{\chi_h} d\chi \, \frac{W_\kappa^i(\chi)\, W_\kappa^j(\chi)}{\chi^2} \, P_\mathrm{NL}\!\left(\frac{\ell + 1/2}{\chi}, z(\chi)\right),
\end{align}
where $\chi$ is the comoving radial distance, $z(\chi)$ is the redshift at comoving distance $\chi$, and $\chi_h$ is the comoving horizon distance. The 3D nonlinear matter power spectrum, $P_\mathrm{NL}$, is computed using the Takahashi halofit fitting formula \cite{takahashi}, and for the intrinsic alignment we consider the Non-Linear Alignment (NLA) model \cite{ia_nla_1,ia_nla_2,ia_nla_3}. 

The lens efficiency kernel in the $i$-th tomographic bin in Equation \eqref{eq:shear_spectra} is given by \cite{cov_modeling_des_y3}:
\begin{align}\label{eq:lensing_efficiency}
    W_\kappa^i(\chi) = \frac{3 H_0^2 \Omega_m \chi}{2 a(\chi)} \int_\chi^{\chi_h} d\chi' \left(\frac{\chi' - \chi}{\chi}\right) n_\kappa^i(z(\chi')) \frac{dz}{d\chi'},
\end{align}
where $H_0$ is the Hubble constant, $\Omega_m$ is the matter density, and $a(\chi)$ is the scale factor at comoving distance $\chi$. The source galaxy redshift distribution in tomographic bin $i$ is denoted by $n_\kappa^i(z)$ with uncertainties in its shape parameterized through the mean-shift $\Delta_z^i$ or the amplitude of the PCs $u_i$'s. 

Starting from a chosen redshift distribution $n(z)$ shown in Figure~\ref{fig:roman_scenarios}, together with the fiducial cosmology listed in Table~\ref{tab:fiducial_cosmology} and an assumed covariance matrix, we compute the posterior distribution $P(\boldsymbol{\theta}|\mathbf{D})$ using Bayes’ theorem $P(\boldsymbol{\theta}|\mathbf{D})\propto\mathcal{L}(\mathbf{D}|\boldsymbol{\theta})P(\boldsymbol{\theta})$ where $P(\boldsymbol{\theta})$ is the prior and $\mathcal{L}(\mathbf{D}|\boldsymbol{\theta})$ is the likelihood function, given by
\begin{align}
    &L\left(\mathbf{D} \,|\, \mathbf{\theta}, \hat{\mathbf{n}}\right) \propto 
    \exp\Bigg[-\frac{1}{2}\left(\mathbf{D}-\mathbf{M}\right)^\mathrm{T} \mathbf{C}^{-1} \left(\mathbf{D}-\mathbf{M}\right)\Bigg],\label{eq:likelihood}\\
    &\chi^2 \equiv (\mathbf{D}-\mathbf{M})^\mathrm{T} \mathbf{C}^{-1} (\mathbf{D}-\mathbf{M}),\label{eq:chi2_full}
\end{align}
where the synthetic cosmic shear data vector is denoted by $\mathbf{D}$, and the model prediction by $\mathbf{M}(\boldsymbol{\theta})$. The parameter vector is decomposed as $\boldsymbol{\theta} \rightarrow {\theta, \mathbf{n}}$ to distinguish the cosmological and nuisance parameters $\theta$ from those describing the redshift distribution $\mathbf{n}$, either from the mean-shift or the PCA method. We use the notation of full $\chi^2$ as defined in Equation \eqref{eq:chi2_full}, which should not be confused with the comoving distance $\chi$ in Equation \eqref{eq:lensing_efficiency}. Reference \cite{pca_method} presents a linearized approximation to the $\chi^2$. This approximation leverages the fact that the most part of the cosmological information is contained in the leading PCs due the optimization (or weights) from the Fisher matrix in Equation \eqref{eq:fisher_matrix}. We adopt the formalism of Ref.~\cite{pca_method} to identify the leading weighted PCs for Roman, while using \texttt{CoCoA} to evaluate the full $\chi^2$. In the next section, we compare this linearized approach with the full $\chi^2$ evaluation comparing the weighted PCA with the unweighted PCA.

\section{Results}\label{sec:results}
In this section, we present the results of our PCA-based parameterization of uncertainties in the redshift-distribution shapes and compare them with the standard mean-shift approach. For both PCA and mean shift, we first select one Roman scenario from Fig.~\ref{fig:roman_scenarios} to define the fiducial redshift distribution and adopt the fiducial cosmological parameters in Table \ref{tab:fiducial_cosmology} to construct the synthetic cosmic shear data vector. The choice of a specific realization \( r \) of the redshift distribution, \( \mathbf{n}^r(z) \), among the one million available is determined by the chi-squared value \( \chi^2_{0,r} \) evaluated at the fiducial parameters. Here, the subscript ``0'' denotes the case with no principal components, since \( \chi^2 \) depends on the number of PCs, i.e., \( \chi^2_0 \equiv \chi^2(0~\mathrm{PC}) \). The mean-shift method requires only this setup, after which nine nuisance parameters $\Delta_z^i$ are varied in the MCMC runs. The PCA-based method introduces two additional steps: selecting a Roman scenario to train the PC modes ($\mathbf{U}_i$) and choosing how many PC amplitudes ($u_i$) to vary in the MCMC analyses. As a result, the PCA workflow involves two independent scenario choices-one for defining the fiducial simulated cosmic shear data vector and another for generating the PC basis. When these coincide, we refer to the case as ``non-mixing''; when they differ, we refer to it as ``mixing''.

\begin{figure}[h]
    \centering
    \includegraphics[width=1\linewidth]{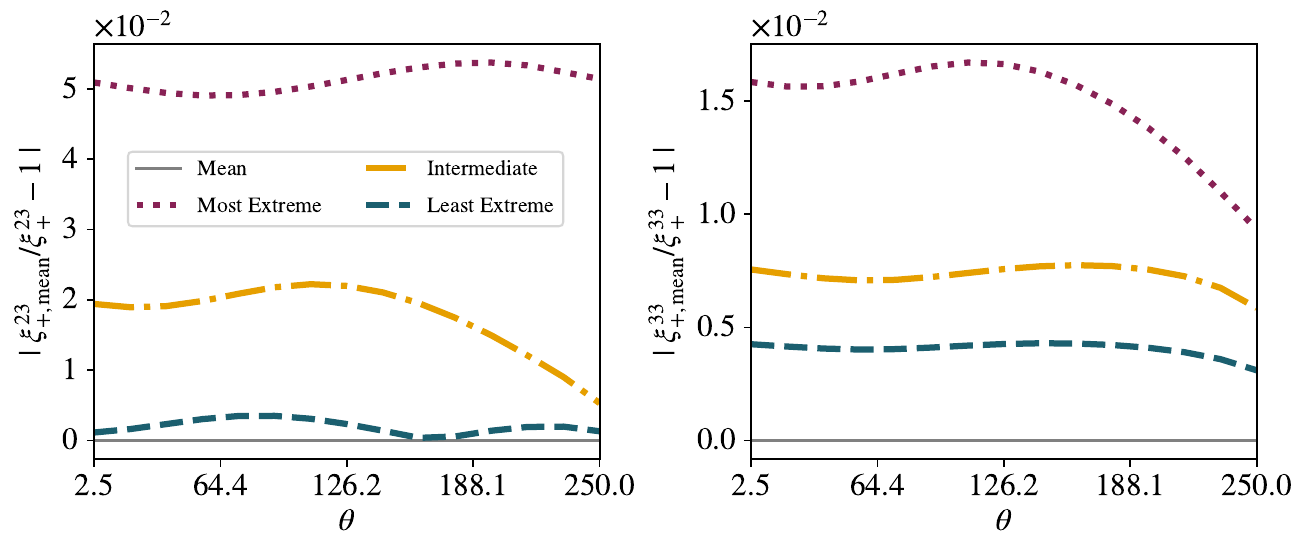}
    \caption{The relative difference of the cosmic shear two-point function $\xi_+^{ij}(\theta)$ for the bins $i=2,j=3$, and $i=3,j=3$, as function of the angular binning, $\theta$, in three different cases of the redshift distribution: ``least extreme'' (green dashed line), ``intermediate'' (orange dashed dotted), and ``most extreme'' (dotted purple). The gray line at 0 corresponds to the $\xi_+^{ij}(\theta)$ for mean redshift distribution.}
    \label{fig:least_intermediate_most_extreme_xi}
\end{figure}

To evaluate how effectively the PCA and mean-shift methods mitigate biases, we follow the definitions in \cite{baryon_pca_wl_2}, with a slight adjustment to the error estimate: the bias is defined as the difference between the fiducial value and the best-fit, and the error is taken as half of the $2\sigma$ width of the 1D marginalized constraint on the parameter of interest. In addition, we present the resulting 2D marginalized posterior distributions for $\Omega_m$ and $S_8$. 

\subsection{Baseline results of PCA photo-z mitigation for Roman}
\begin{figure}[h]
    \centering
    \includegraphics[width=1\linewidth]{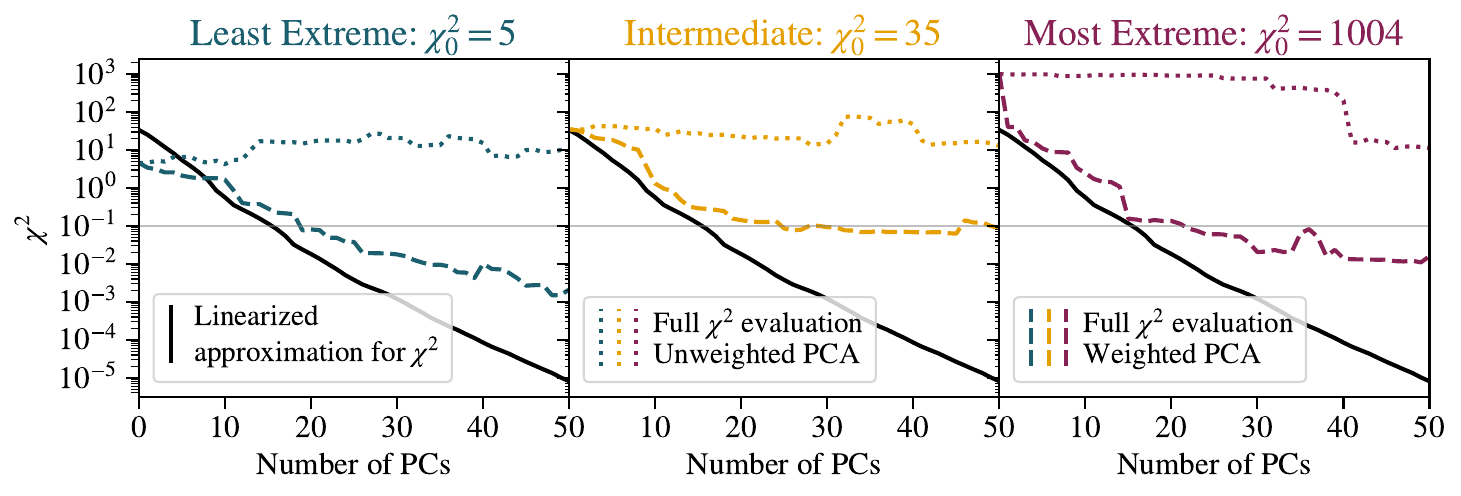}
    \caption{The $\chi^2$ as a function of the number of PCs. The black solid line shows the linearized approximation, while the dotted and dashed lines show the full $\chi^2$ evaluations using the unweighted and weighted PCA, respectively. The titles ``least extreme'' ($\chi^2_0 = 5$), ``intermediate'' ($\chi^2_0 = 35$), and ``most extreme'' ($\chi^2_0 = 1004$) refer to the value of $\chi^2$ at zero PCs. The gray horizontal line marks $\chi^2 = 0.1$ and indicates the number of PCs required to remain below this threshold: approximately 20 PCs for the least and most extreme realizations and about 25 PCs for the intermediate case. Beyond $\sim$20 PCs, adding more modes does not yield a significant gain in information, as also demonstrated in Fig.~\ref{fig:violinplot}.}
    \label{fig:comparison_least_intermediate_most_extreme}
\end{figure}

We start with the ``non-mixing'' choice of Roman scenarios, adopting the same W1-D1 to build both the synthetic cosmic shear data vector and the PCs. We then compute a single likelihood evaluation for several realizations $r$ to obtain the initial chi-squared values, $\chi_{0,r}^2$. Each evaluation with \texttt{CoCoA} takes about 0.18 seconds, allowing us to quickly estimate the goodness of fit. The realization indices $r$ are then ranked in order of increasing $\chi_{0,r}^2$, and three representative cases are selected: the ``least extreme'' realization, defined by $\chi_{0,\mathrm{least}}^2$, with an arbitrarily low value; the ``most extreme'' realization, $\chi_{0,\mathrm{most}}^2$, with an arbitrarily high value; and the ``intermediate'' realization, $\chi_{0,\mathrm{inter}}^2$, with a value between the previous two. The impact of selecting representative realizations on $\xi_\pm^{ij}(\theta)$ and on $\chi^2$ as a function of the number of PCs is shown in Figures~\ref{fig:least_intermediate_most_extreme_xi} and \ref{fig:comparison_least_intermediate_most_extreme}, respectively. The results correspond to the least, intermediate, and most extreme $\mathbf{n}(z)$ realizations, with initial reduced chi-square values $\chi^2_{0,r}=$ 5, 35, and 1004, respectively. In Fig.~\ref{fig:least_intermediate_most_extreme_xi} we show how each of these three cases affects $\xi_\pm^{ij}(\theta; \mathbf{n}_r)$ relative to $\xi_\pm^{ij}(\theta;\bar{\mathbf{n}})$. We have choose the tomographic bins combinations $(i,j)=(2,3)$ and $(3,3)$ to preserve the sense ``most'' $>$ ``intermediate'' $>$ ``least'' in the data vector compared to the mean, but one should notice that $\chi^2$ depends on the covariance matrix, thus a realization can exhibit a larger deviation at specific angular scales (or for specific tomographic bin combinations) while still having a smaller overall $\chi^2$.

In Fig.~\ref{fig:comparison_least_intermediate_most_extreme} we show the $\chi^2$ values obtained as an increasing number of PCs is included in the analysis, following Equation \eqref{eq:n_in_e_base}, for the least, intermediate, and most extreme cases. The PCA-based approach implemented in \texttt{CoCoA}, parameterize the photo-z uncertainty through $\bar{\mathbf{n}}(z) + \sum_i u_i \mathbf{U}_i(z)$, and evaluates $\chi^2$ using the full likelihood expression given in Equation \eqref{eq:likelihood}. That said, the work of \cite{pca_method}, which forms the basis of our analysis, provides a linearized approximation for $\chi^2$, given in their Equation 17. We therefore compare the $\chi^2$ evaluated using the full likelihood implemented in \texttt{CoCoA}, with the corresponding linearized expression of \cite{pca_method}. We highlight, however, that the comparison of these two cases we do in the following are not meant to indicate which one is ``better'', but to provides a consistency check between our numerical implementation in \texttt{CoCoA} (as part of the Roman HLIS Cosmology PIT analysis pipeline) with the the analytical linearized approximation of the precursor work \cite{pca_method}. 

Fig.~\ref{fig:comparison_least_intermediate_most_extreme} shows that the linearized approximation to $\chi^2$ decays more rapidly than the full $\chi^2$ evaluations. This behavior is expected: the linearized case depends only on the sum of the eigenvalues from the eigensystem decomposition optimized via the Fisher matrix in Equation \eqref{eq:fisher_matrix} (see \cite{pca_method} for further details). In contrast, the weighted and unweighted full $\chi^2$ evaluation from Equation \eqref{eq:chi2_full} estimated with \texttt{CoCoA} exhibit a nontrivial dependence on the PC modes of Equation \eqref{eq:n_in_e_base}. Nevertheless, the full $\chi^2$ evaluation for the weighted PCA case exhibits an overall decay similar to that of the linearized approximation up to 20-30 PCs depending of the scenario, demonstrating the consistency of our numerical implementation in \texttt{CoCoA}. By contrast, the weighted case outperforms the unweighted one, as expected, since the Fisher matrix weights the PC modes by emphasizing directions in parameter space that are better constrained, thereby improving the stability and accuracy of the reconstructed observables. Although this behavior may seem obvious, we include the unweighted $\chi^2$ evaluation to underscore the critical role of the Fisher matrix in optimizing the $\chi^2$ computation.

\begin{figure}[h]
    \centering
    \includegraphics[width=\linewidth]{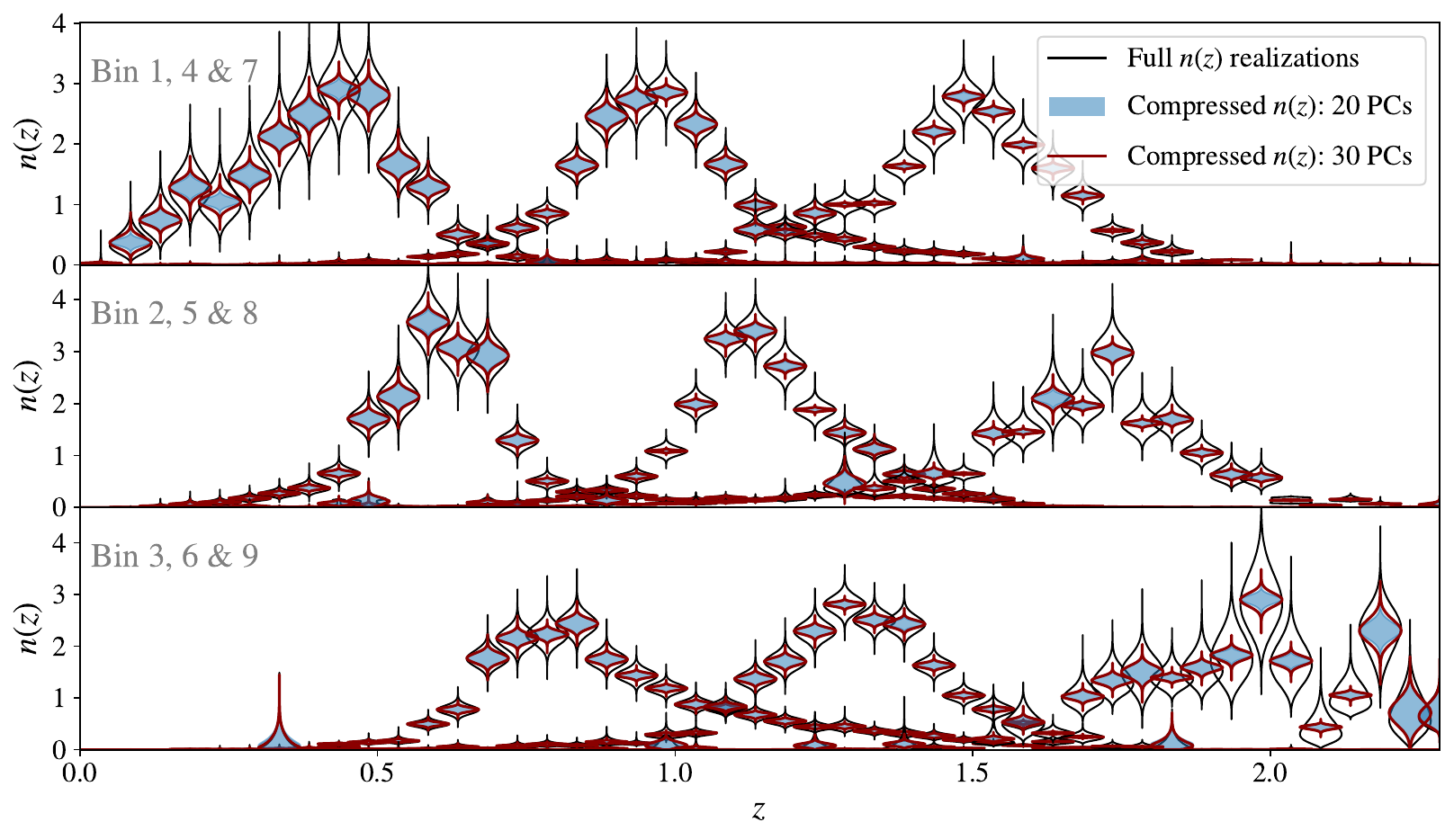}
    \caption{Violin plots of the redshift distribution $n(z)$ for 9 tomographic bins and 46 redshift sub-bins. The black contours show the full set of $10^6$ $n(z)$ realizations for the Roman W1--D1 scenario. The blue and dark-red contours correspond to compressed representations using 20 and 30 principal components, respectively. The 30-mode case does not provide additional information compared to the 20-mode case, indicating that 20 principal components are sufficient.}
    \label{fig:violinplot}
\end{figure}

Although $\chi^2_{0,\mathrm{least}}\ll\chi^2_{0,\mathrm{most}}$, we can see from Fig.~\ref{fig:comparison_least_intermediate_most_extreme} that this does not guarantee faster convergence of the ``least extreme'' case compared to the ``most extreme'' one. The threshold of $\chi^2\leq0.1$ is achieved nearly at 20 PCs for each extreme scenario. This result suggests that $\chi^2_0$ serves only to set a higher bias, not to guarantee a faster drop in $\chi^2$ when more PCs are included. This makes sense: any $\mathbf{n}$ shares the same space spanned by the PC modes, which are ordered by the amount of variance they explain in the realizations, from highest to lowest. If a given $\mathbf{n}(z)$ is the ``least extreme'', $\mathcal{O}(\chi_0^2)\leq1$, then we have to add higher order PCs to see significant drop in the $\chi^2$. If $\mathbf{n}(z)$ is the ``most extreme'' scenario, $\mathcal{O}(\chi_0^2)\geq 10^3$, then just adding the first PCs causes a faster drop in $\chi^2$. Once the most relevant modes are included, $\chi^2$ starts converging slowly towards zero because the highest modes do not add significant information. In Fig.~\ref{fig:comparison_least_intermediate_most_extreme}, from 0-10 PCs, the ``least extreme'' drops from $\chi_0^2=5$, to $\chi^2=1.6$, while the ``most extreme'' drops three orders of magnitudes, from $\chi_0^2=1004$ to $\chi^2=2.5$. The saturation of information beyond 20 PCs is illustrated in the violin plot of Fig.~\ref{fig:violinplot}. For clarity, each row shows the distribution of $\mathbf{n}(z)$ for every third tomographic bin. The black contours represent the full set of one million $\mathbf{n}(z)$ realizations for the Roman W1-D1 scenario. The shaded blue region shows the reconstruction using the first 20 PCs (out of 414 in total), while adding more PCs does not lead to any visible improvement, as illustrated by the red contours.

We note, however, that we are comparing only three realizations within a landscape of one million. Moreover, because the full likelihood in Equation \eqref{eq:likelihood} is a nonlinear function of the redshift distribution computed numerically with \texttt{CoCoA}, we refrain from generalizing the decay of the full $\chi^2$ evaluations as a function of the number of PC modes; in fact, $\chi^2$ can occasionally rise before decreasing again. Although many factors can affect the overall trend-such as the simulated redshift distributions, the construction of the fiducial cosmic shear data vector, the scale cuts, and the data covariance matrix-it is notable that the weighted PCA in the full $\chi^2$ evaluations tends to approach the linearized approximation discussed by \cite{pca_method}, underscoring the consistency of our implementation in \texttt{CoCoA}.

\subsection{Comparison between PCA and mean-shift uncertainty parameterizations}
\begin{figure}[h]
    \centering
    \includegraphics[width=\linewidth]{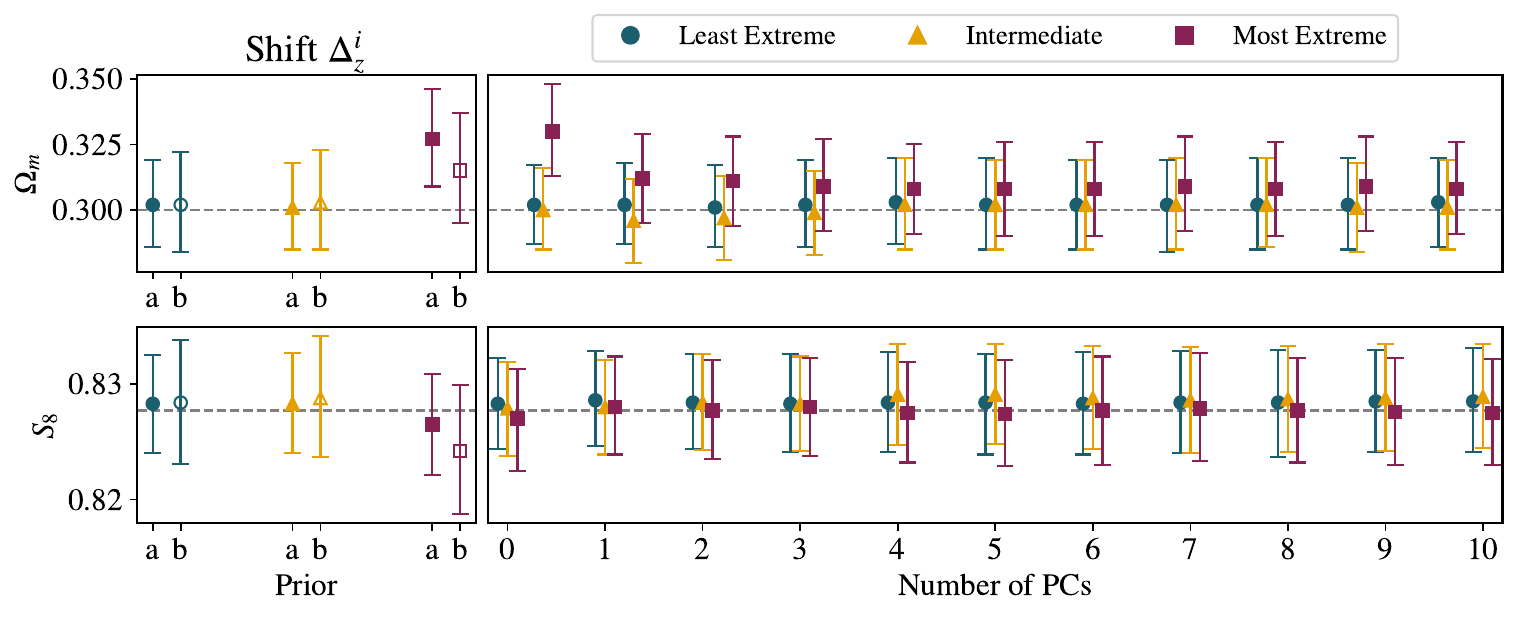}
    \caption{Constraints on $\Omega_m$ and $S_8$ at $2\sigma$ for the mean-shift and PCA methods for photo-$z$ systematic parameterization. 
The former considers nine parameters $\Delta_z^i$, corresponding to the number of tomographic bins, and is analyzed under two priors, ``a'' and ``b'' (see Table~\ref{tab:fiducial_cosmology}). 
The PCA approach is analyzed under the same three extreme scenarios as in Figure~\ref{fig:comparison_least_intermediate_most_extreme}, i.e., synthetic cosmic shear data vectors generated with a specific $n(z)$ such that, when computing the chi-squared at zero PCs, they yield one of three cases: 
$\chi_0^2 = 5$ (``least extreme''), $\chi_0^2 = 35$ (``intermediate''), and $\chi_0^2 = 1004$ (``most extreme'').}
    \label{fig:mitigation_least_intermediate_most_extreme}
\end{figure}

We use the least, intermediate, and most extreme cases from Fig.~\ref{fig:comparison_least_intermediate_most_extreme} to run Markov Chain Monte Carlo, assessing the bias-mitigation performance of the weighted PCA approach versus the standard mean-shift method. Fig.~\ref{fig:mitigation_least_intermediate_most_extreme} presents the 1D constraints on $\Omega_m$ and $S_8$ with $2\sigma$ error bars with representative values shown in Table \ref{tab:least_most_extreme_summary}. The left column shows how these constraints vary with the number of PCs. For the least extreme and intermediate scenarios, including additional PCs does not improve the constraints on $\Omega_m$ and $S_8$. This is expected, as the initial $\chi_\mathrm{ini}^2$ values-corresponding to no mitigation (\# of PCs = 0)-are already low: $5$ and $35$, respectively. The same conclusion applies to the mean-shift approach (right column). Notice, however, that widening the prior on $\Delta_z^i$ by nearly a factor of 3.33 (from $\mathrm{Gauss}(0, 0.003)$ to $\mathrm{Gauss}(0, 0.01)$) improves the accuracy, but warrants an increase in the error bars of $\Omega_m$ and $S_8$. 

\begin{table}[h]
\centering
\caption{Summary of the marginalized constraints on $\Omega_m~(\Omega_m^\mathrm{fid})$ and $S_8~(S_8^\mathrm{fid})$ for the least- and most-extreme redshift realizations. For the PCA mitigation scheme, we report results for the cases with 0, 5, and 10 retained PCs. For the mean-shift mitigation scheme, results are shown for priors ``a'' and ``b''.}
\vspace{0.1cm}
\label{tab:least_most_extreme_summary}
\begin{tabular}{ccccc}
\toprule
Method & Realization & \# PCs / Prior & $\Omega_m~(0.3)$ & $S_8~(0.8277)$ \\
\toprule
\multirow{6}{*}{PCA}
& \multirow{3}{*}{Least extreme}
& 0  & $0.302^{+0.015}_{-0.015}$ & $0.8283^{+0.0040}_{-0.0039}$ \\
& & 5  & $0.302^{+0.018}_{-0.017}$ & $0.8284^{+0.0042}_{-0.0045}$ \\
& & 10  & $0.303^{+0.017}_{-0.017}$ & $0.8285^{+0.0046}_{-0.0044}$ \\
\cmidrule(lr){2-5}
& \multirow{3}{*}{Most extreme}
& 0  & $0.330^{+0.018}_{-0.017}$ & $0.8270^{+0.0043}_{-0.0045}$ \\
& & 5  & $0.308^{+0.018}_{-0.018}$ & $0.8274^{+0.0047}_{-0.0045}$ \\
& & 10  & $0.308^{+0.018}_{-0.017}$ & $0.8275^{+0.0047}_{-0.0045}$ \\
\midrule
\multirow{4}{*}{Mean-shift}
& \multirow{2}{*}{Least extreme}
& a & $0.302^{+0.017}_{-0.016}$ & $0.8283^{+0.0042}_{-0.0043}$ \\
& & b & $0.302^{+0.020}_{-0.018}$ & $0.8284^{+0.0054}_{-0.0053}$ \\
\cmidrule(lr){2-5}
& \multirow{2}{*}{Most extreme}
& a & $0.327^{+0.019}_{-0.018}$ & $0.8265^{+0.0044}_{-0.0044}$ \\
& & b & $0.315^{+0.022}_{-0.020}$ & $0.8242^{+0.0057}_{-0.0055}$ \\
\bottomrule
\end{tabular}
\end{table}

The most stringent test for the PCA and mean-shift schemes for mitigating photo-z systematics occurs in the most extreme scenario. We observe by Fig.~\ref{fig:mitigation_least_intermediate_most_extreme} a clear improvement in the bias mitigation of $\Omega_m$ as additional PCs are included. With just one PC, $\Omega_m$ falls within $2\sigma$ of the fiducial value of 0.3, whereas the mean-shift approach does not reach the same level of agreement for the tightest prior, but does if the Gaussian width is increased to 0.01 as indicated by the hollow square in Fig.~\ref{fig:mitigation_least_intermediate_most_extreme}.   

\subsection{The impact of Roman deep field design on \texorpdfstring{$n(z)$}{} estimation}\label{sec:results2}
The results presented so far consider the situation where the PC modes were trained on the same Roman observing scenario (i.e., W1-D1) that was used to generate the fiducial synthetic data vector. In this section, we explore training and testing with different Roman observing scenarios (see Table \ref{tab:roman_scenarios}; e.g., using W1-D2 for training and W1-D1 for testing). 

The aim of this exploration is to assess how the weighted PCA model performs when applied to data that differ from the sample used to construct the PC basis. This test probes the internal consistency of the method across different simulated survey configurations. However, it should not be interpreted as a demonstration of performance on real data, as additional sources of mismatch between simulations and observations are not included in this analysis. 

As shown in the lower-triangle panel of Fig.~\ref{fig:alphas_distributions_double}, using PCs and synthetic $\xi_\pm$ data vectors derived from different Roman scenarios (the mixing case) introduces a slight skewness in the distribution of the PC amplitudes, though we still adopt a Gaussian (0, 1) prior in the MCMC runs. As in the non-mixing case, we must choose a redshift distribution to construct the fiducial synthetic $\xi_\pm$ data vector: for the non-mixing analysis we use the Least Extreme, Intermediate, and Most Extreme cases, while for the mixing case we use the mean of the one million redshift distributions. Table \ref{tab:mixing_scenarios} summarizes the analysis choices for the mixed-scenario setup. In this configuration, the PCs are computed from the Roman deep-field scenarios D2, D3, and D4 (i.e., with magnitude errors, and see Table \ref{tab:roman_scenarios}), while the $\xi_\pm$ data vector is generated from the idealized deep-field W123-D1, i.e., error free scenarios.

\begin{table}[H]
    \centering
    \begin{tabular}{|c|ccc|}
         \hline
         PC training (PCs) & W1-D234 & W2-D234 & W3-D24 \\
         \hline
         Fiducial $\xi_\pm$ data vector (DV) & W1-D1 & W2-D1 & W3-D1\\
         \hline
    \end{tabular}
    \caption{Combinations of Roman observing scenarios for the simulated analyses.}
    \label{tab:mixing_scenarios}
\end{table}

\begin{figure}[h]
    \centering
    \includegraphics[width=1\linewidth]{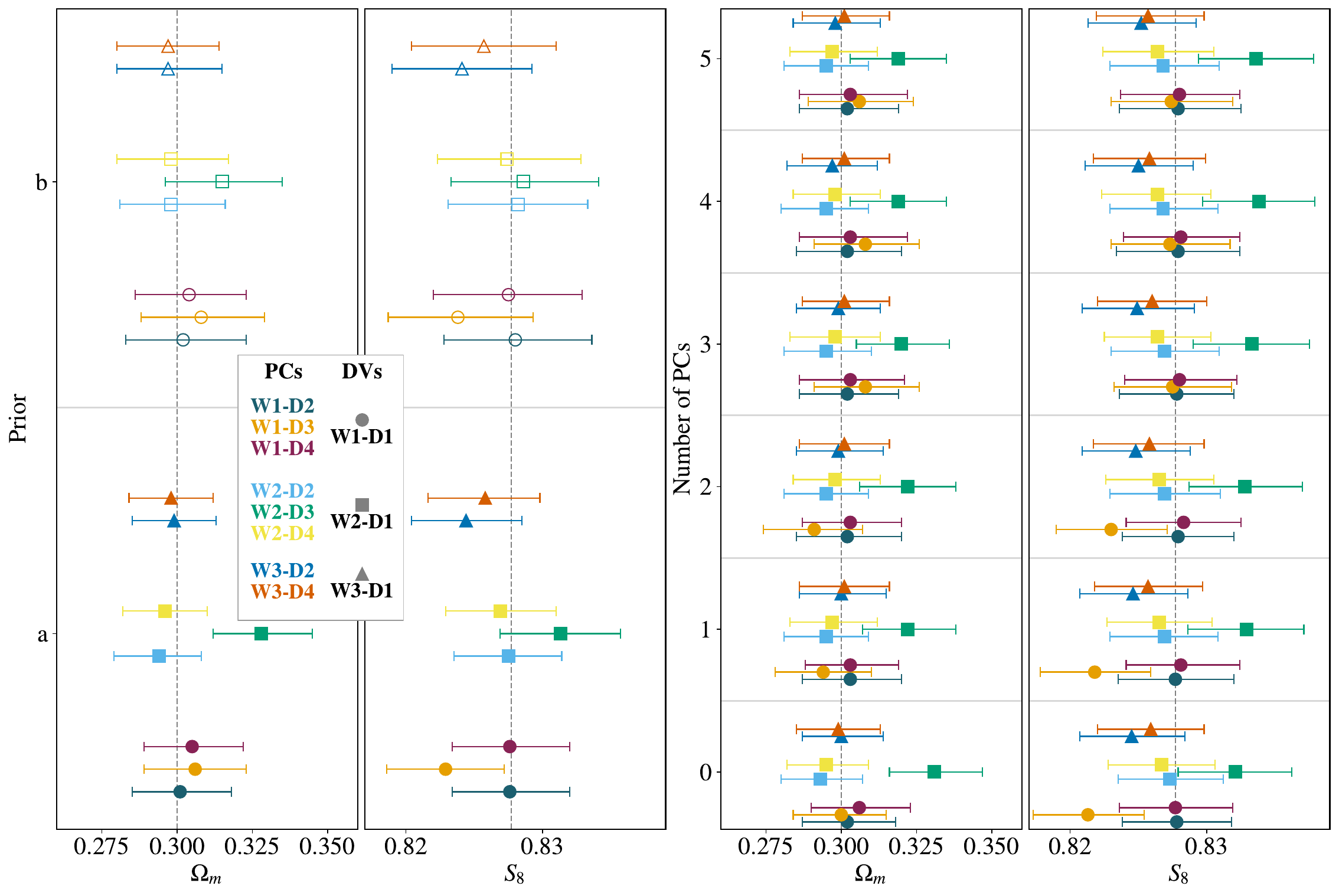}
    \caption{Constraints on $\Omega_m$ and $S_8$ at the $2\sigma$ confidence level. From left to right; the first two columns are constraints using the mean-shift approach for a tight prior (``a'') and wide prior (``b''). The last two columns, use the PCA method up to five PCs. The markers (circles, squares, and triangles) indicate the Roman scenario used to generate the synthetic $\xi_\pm$ data vector from a particular mean redshift distribution; circles correspond to $\bar{n}(z)$ from W1-D1, squares from W2-D1, and triangles from W3-D1. For the PCA method, the different colors denote the Roman scenarios employed to construct the PCs.}
    \label{fig:mitigation_mixing_scenarios}
\end{figure}

In Fig.~\ref{fig:mitigation_mixing_scenarios},
we show the $2\sigma$ constraints on $\Omega_m$ and $S_8$ for both the PCA and mean-shift mitigation schemes. For the combinations (DV: W1-D1, PCs: W1-D234), (DV: W2-D1, PCs: W2-D234), and (DV: W3-D1, PCs: W3-D24), the two methods remain consistent with each other and yield comparable constraints for most part of these combinations. However, both approaches are challenged in addressing the bias in two specific situations. The first case is indicated by the orange circles that considers the combination (DV: W1-D1, PCs: W1-D3). The PCA method mitigate the bias in $S_8$ once three or more PC modes are added. The mean-shift approach does not recover the fiducial $S_8$ value at $2\sigma$ for the tighter prior ``a'', unless the broader one ``b'' is assumed.

The second challenge case is illustrated by the light-green squares, corresponding to the combination (DV: W2-D1, PCs: W2-D3), which exhibits a strong bias. The mean-shift approach fails to mitigate the bias in $\Omega_m$ under the more restrictive prior ``a'', except when the broader prior ``b'' is adopted, which warrants increased error bars. Conversely, PCs constructed from W2-D3---even when including up to five modes---do not meaningfully improve constraints on the higher-order moments of the redshift distribution for the W2-D1 scenario.

\begin{figure}[h]
    \centering
    \includegraphics[width=1\linewidth]{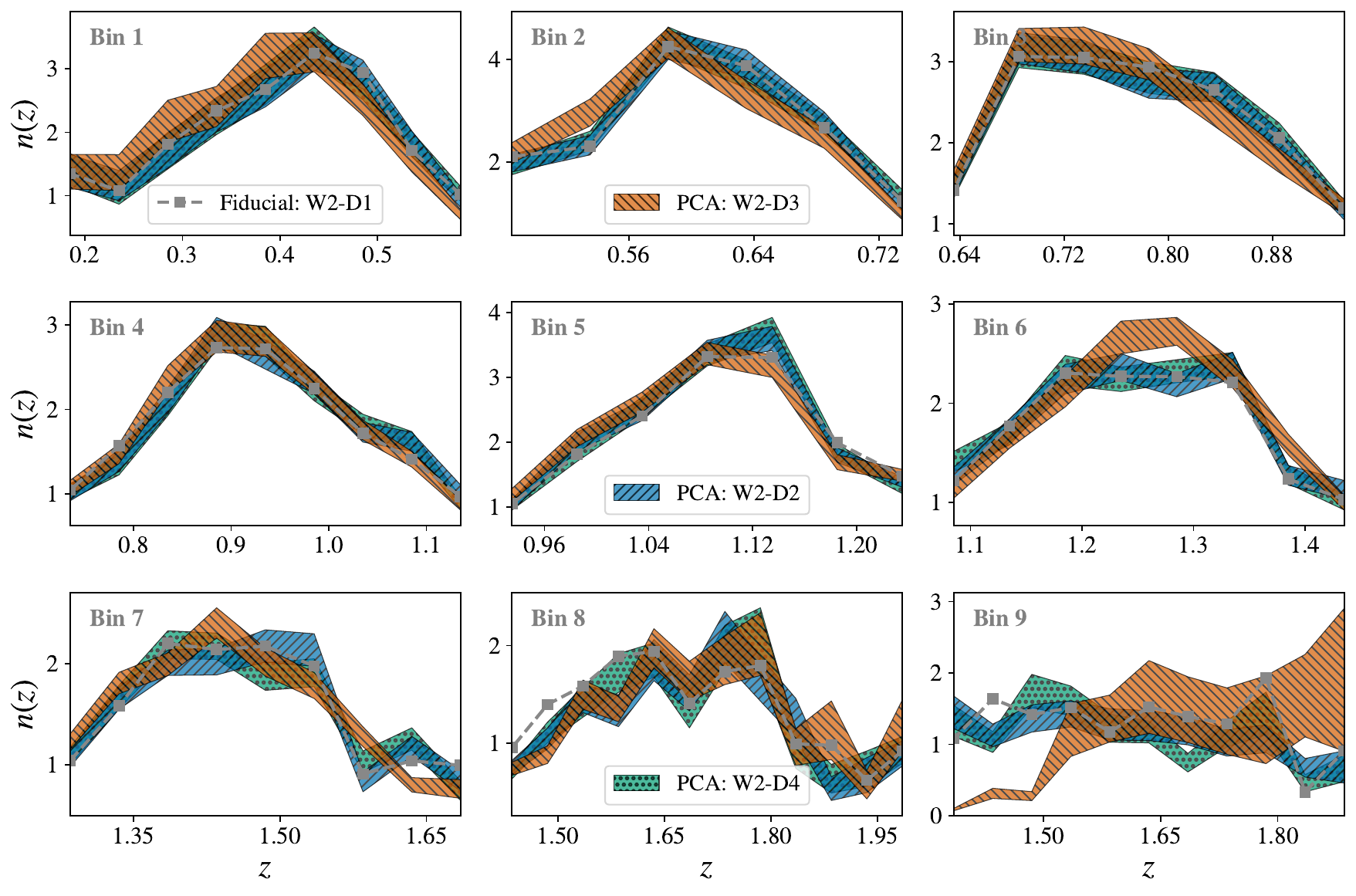}
    \caption{Redshift distributions for the fiducial scenario W2-D1 (the mean $n(z)$) and the three other scenarios used to build the PC basis: W2-D2 and W2-D4, with areas of $20~\deg^2$, and W2-D3 with an area of $40~\deg^2$. The shaded bands correspond to the 68\% intervals. The redshift range was truncated on both sides to show more clearly the relevant range around the peaks of the redshift distributions.}
    \label{fig:referee_major2_68CL_9bins_used}
\end{figure}

To understand the failure case D3, in Fig.~\ref{fig:referee_major2_68CL_9bins_used} we show the $n(z)$ distributions in the nine tomographic bins for the specific ``chunk'' corresponding to the three colored ``squares'' in Fig.~\ref{fig:mitigation_mixing_scenarios}. The W2-D2 and W2-D4 scenarios remain broadly consistent with the fiducial W2-D1 distribution across most tomographic bins. In contrast, the W2-D3 scenario shows several redshift regions where the $68\%$ interval of the calibrated $n(z)$ realizations does not fully enclose the fiducial W2-D1 mean $n(z)$, which directly affects the performance of the PCA and mean-shift methods.

The PCA basis is constructed from the ensemble of W2-D3 realizations, the dominant PCA modes describe fluctuations around this shifted ensemble rather than around the fiducial W2-D1 distribution itself. Consequently, increasing the number of retained PCs does not necessarily drive the inferred cosmological constraints back toward the fiducial cosmology, as seen in the $\Omega_m$ and $S_8$ panels of Fig.~12 in the paper. We highlight, however, that the same qualitative behavior is also observed for the mean-shift calibration method (prior ``a''), where the W2-D3 scenario remains the only case that does not recover the fiducial $\Omega_\mathrm{m}$ within the quoted uncertainties. 

\section{Conclusion}\label{sec:conclusion}
Designing a balance of imaging at different depths is critical for the goals of the HLWAS; probe the growth of the large scale structure and the expansion of the Universe with weak gravitational lensing and galaxy clustering, as well enabling several of others general astrophysics science. The nominal (in-guide) HLWAS consider a deep tier of $19.2\deg^2$ providing crucial calibration data and a wide tier of $2700\deg^2$ in H-band only (see \cite{rotac_report} for further details). Specifically, the tomographic analysis of weak gravitational lensing is a powerful probe of the late Universe; however, uncertainties in the redshift distribution can bias the cosmological analysis and ultimately impede our ability to distinguish between dark energy models. In this work, we explore two ways of modeling redshift-distribution uncertainties for the Roman Space Telescope: the standard mean-shift parameterization and the newer PCA-based approach proposed in \cite{pca_method}, which was adopted in the DES Y6 analyses by \cite{des_y6_pca_redshift_calib_source} and \cite{des_y6_pca_redshift_calib_lens}. We implemented the PCA-based approach of \cite{pca_method} in the \texttt{CoCoA} code, which is part of the Roman HLIS Cosmology PIT analysis pipeline. To test our PCA-based implementation in \texttt{CoCoA} for modeling uncertainties in the redshift distribution and to assess its impact on cosmological parameter inference, we used the \texttt{Cardinal} to generate a mock galaxy catalogs and the SOMPZ methodology for bin assignment and draw millions of redshift-distributions. The uncertainty of the redshift distributions are to some extent stemmed from the assumptions of the wide- and deep-tiers configurations, such as the magnitude error, exposure times, filters, area and limiting magnitude in the H band. As a result, the choice of wide- and deep-tier configurations affects the uncertainty in the estimated redshift distributions and propagates into biases in the inferred cosmological parameters. 

In this work, we therefore explore a wide range of scenarios: three wide tiers, labeled W1, W2, and W3, and four deep tiers, labeled D1, D2, D3, and D4. We combine these tiers to form a set of eleven possible Roman HLWAS survey scenarios: W1-D1, W1-D2, W1-D3, W1-D4, W2-D1, W2-D2, W2-D3, W2-D4, W3-D1, W3-D2, and W3-D4. We do not include the W3-D3 configuration, as it does not probe an additional calibration regime beyond those already explored and is therefore outside the scope of our targeted stress test of enlarged deep-field areas for the baseline wide tiers (W1 and W2). In comparison with the nominal HLWAS recommended by the ROTAC, i.e., a deep tier covering $19.2\deg^2$ and a wide tier covering $2700\deg^2$, the deep tiers D1, D2, and D4 cover the same area of $20\deg^2$, while the deep tier D3 covers $40\deg^2$. Therefore, wide tiers paired with D3 deviate from the ROTAC recommendations. However, we include these non-recommended scenarios to stress test our PCA implementation in the \texttt{CoCoA} framework, as it is an essential component of the HLIS PIT pipeline. We then perform multiple MCMC analyses using both mitigation schemes-the mean-shift and PCA approaches-and compare their bias mitigation and constraining power on the cosmological parameters $\Omega_m$ and $S_8$.

We found that when the $\xi_\pm$ data vector (generated from an $\mathbf{n}(z)$ realization) is not strongly miscalibrated-i.e., has a low initial value of $\chi^2_0\sim O(10)$-relative to our best estimate of the redshift distribution (the mean $\bar{\mathbf{n}}(z)$ across realizations), both the mean-shift and PCA-based methods provide comparable constraints on the parameters $\Omega_m$ and $S_8$, a trend consistent with the findings of \cite{des_y6_pca_redshift_calib_source} in the DES Y6 cosmic-shear analysis. However, meaningful differences between the two approaches emerge when we introduce moderate ($\chi^2_0\sim O(100)$) to strong ($\chi^2_0\sim O(1000)$) miscalibrations. In these regimes, using a tight Gaussian prior (mean = 0, std = 0.003) on the $\Delta_z^i$ parameters fails to mitigate the resulting biases in $\Omega_m$ and $S_8$. Relaxing the prior to a wider Gaussian (mean = 0, std = 0.01) helps reduce these biases, but at the cost of mildly larger error bars (an increase of $\sim13.5\%$ in the $\Omega_\mathrm{m}$ uncertainty). In the most extreme scenario, for the PCA-based approach (see Fig.~\ref{fig:mitigation_least_intermediate_most_extreme}), using 5 PCs mitigates the bias in $\Omega_m$, and adding additional PCs does not significantly improve the constraints, whereas the 9 $\Delta_z^i$ parameters yields slightly weaker constraints under the wider prior. It is also worth noting that we adopt a relatively broad Gaussian prior (mean = 0, std = 1) on the PC amplitudes, $u_i$. Although this prior is not directly comparable to the mean-shift priors (std = 0.003 and 0.01), the PCA-based method still yields meaningful constraints on the cosmological parameters. A more informative prior on these amplitudes would be expected to further tighten the constraints on $\Omega_m$ and $S_8$. Although the PCA-based approach analyzed here performs comparably to the standard mean-shift parameterization in mitigating biases--yielding consistent cosmological constraints in the least and intermediate scenarios--we emphasize that these results are obtained under deliberately extreme configurations designed to stress-test the $n(z)$ PCA implementation in \texttt{CoCoA}. In the most extreme case, differences between the two approaches become more apparent, reflecting the large initial bias introduced at zero PCs. Nonetheless, our results indicate that the PCA-based method provides a flexible and potentially promising framework for modeling redshift-distribution uncertainties and may offer modest improvements in constraining power in certain regimes.

In summary, it is worthy of attention that every scenario with wide- and deep-tier close to the nominal HLWAS, i.e., D1, D2 and D4 , possess constraints on $\Omega_m$ and $S_8$ within $2\sigma$ confidence level of its respective fiducial values via PCA-based approach or the standard mean-shift method. However, the scenarios with deep tier D3 are the most challenging in mitigating biases on these same cosmological parameters; for the scenario W1-D3, the bias on $S_8$ is mitigated with three PC modes, but one needs to relax the prior condition with the mean-shift approach in order to attain the mitigate the bias. On the other hand, the scenario W2-D3 is the most challenge, indeed the PCA-method does not mitigate in either parameter, $\Omega_m$ neither $S_8$, and the mean shift only mitigate the bias on $\Omega_m$ with the relaxed prior, that implies slightly increased error bars  (check Fig.~\ref{fig:mitigation_mixing_scenarios} for further details). Essentially, the difficulty in mitigate bias on $\Omega_m$ and $S_8$ with the PC modes derived from W2-D3 is due to the low variability in the W2-D3 realizations due to the larger deep-field area ($40\deg^2$), which reduces cosmic variance and limits the informativeness of the PCs. The results presented here demonstrates that the PCA-based approach for modeling uncertainties in redshift distribution performs well in most cases, and a part of the instances where its effectiveness is limited, as we highlighted above, the PCA-based framework showed up as viable and possible modeling of photo-z uncertainties to be exported to others imaging surveys such as LSST and Euclid. 
 

Future analyses using the PCA-based technique could incorporate new simulated Roman redshift distributions from \texttt{Cardinal} and the SOMPZ methodology, explore models beyond $\Lambda$CDM-such as dynamical dark energy models (e.g., $w_0w_a$CDM)-extend the analysis to the $3\times2$ correlation functions, and employ emulators to reduce computational costs when deriving cosmological parameter constraints.

\acknowledgments
We thank Gary M.~Bernstein for insightful discussions. DHFS is supported by the Jet Propulsion Laboratory and the California Institute of Technology. 
This research was carried out at the Jet Propulsion Laboratory, California Institute of Technology, and was sponsored by the National Aeronautics and Space Administration (80NM0018D0004).
This work was supported by the NASA ROSES grant 22-ROMAN11-0011, contract number 80NM0024F0012, via a JPL subaward. The High Performance Computing resources used in this investigation were provided by funding from the JPL Enterprise Technology, Strategy, and Cybersecurity Directorate. 

\newpage
\bibliographystyle{JHEP}
\bibliography{nz_pca.bib}

\end{document}